\documentclass[aps,prx,twocolumn,superscriptaddress,floatfix]{revtex4-2}

\usepackage{graphicx}
\usepackage{dcolumn}

\usepackage{bm}
\usepackage{amssymb}
\usepackage{amsmath}
\usepackage{microtype}
\usepackage{xfrac}
\usepackage{physics}
\usepackage{array}
\usepackage[dvipsnames]{xcolor}
\usepackage[colorlinks,plainpages=false,linkcolor=blue,urlcolor=blue,citecolor=blue,pdfpagemode=UseNone,pdfstartview=FitBH]{hyperref}
\usepackage{nicematrix}

\newcommand{\suggest}[1]

\newcommand{\angstrom}{\text{\normalfont\AA}} 
\renewcommand\vec{\mathbf}

\begin{document}
\def\figureautorefname{Fig.}
\def\equationautorefname~#1\null{Eq. (#1)\null}

\title{Single Crystal Diffuse Neutron Scattering Study of the Dipole-Octupole Quantum Spin Ice Candidate \texorpdfstring{Ce$_2$Zr$_2$O$_7$}~: No Apparent Octupolar Correlations Above \textit{T} = 0.05 K}

\author{E.~M.~Smith}
\affiliation{Department of Physics and Astronomy, McMaster University, Hamilton, Ontario L8S 4M1, Canada}

\author{R.~Sch\"{a}fer}
\affiliation{Department of Physics, Boston University, Boston, Massachusetts 02215, USA}

\author{J.~Dudemaine}
\affiliation{D\'epartement de Physique, Universit\'e de Montr\'eal, Montr\'eal, Qu\'ebec, Canada}
\affiliation{Regroupement Qu\'eb\'ecois sur les Mat\'eriaux de Pointe (RQMP)}

\author{B.~Placke}
\affiliation{Max Planck Institute for the Physics of Complex Systems, N\"{o}thnitzer Stra{\ss}e 38, Dresden 01187, Germany}
\affiliation{Rudolf Peierls Centre for Theoretical Physics, University of Oxford, Oxford OX1 3PU, United Kingdom}

\author{B.~Yuan}
\affiliation{Department of Physics and Astronomy, McMaster University, Hamilton, Ontario L8S 4M1, Canada}

\author{Z.~Morgan}
\affiliation{Neutron Scattering Division, Oak Ridge National Laboratory, Oak Ridge, Tennessee 37831, USA}

\author{F.~Ye}
\affiliation{Neutron Scattering Division, Oak Ridge National Laboratory, Oak Ridge, Tennessee 37831, USA}

\author{R.~Moessner}
\affiliation{Max Planck Institute for the Physics of Complex Systems, N\"{o}thnitzer Stra{\ss}e 38, Dresden 01187, Germany}

\author{O.~Benton}
\affiliation{Max Planck Institute for the Physics of Complex Systems, N\"{o}thnitzer Stra{\ss}e 38, Dresden 01187, Germany}
\affiliation{School of Physical and Chemical Sciences, Queen Mary University of London, London, E1 4NS, United Kingdom}

\author{A.~D.~Bianchi}
\affiliation{D\'epartement de Physique, Universit\'e de Montr\'eal, Montr\'eal, Qu\'ebec, Canada}
\affiliation{Regroupement Qu\'eb\'ecois sur les Mat\'eriaux de Pointe (RQMP)}
\affiliation{Institut Courtois, Complexe des sciences, Universit\'e de Montr\'eal, 1375 Ave. Th\'er\`ese-Lavoie-Roux, Montr\'eal, Qu\'ebec H2V 0B3, Canada}

\author{B.~D.~Gaulin}
\affiliation{Department of Physics and Astronomy, McMaster University, Hamilton, Ontario L8S 4M1, Canada}
\affiliation{Brockhouse Institute for Materials Research, McMaster University, Hamilton, Ontario L8S 4M1, Canada}
\affiliation{Canadian Institute for Advanced Research, 661 University Avenue, Toronto, Ontario M5G 1M1, Canada.}

\date{\today}

\begin{abstract} 
The insulating magnetic pyrochlore Ce$_2$Zr$_2$O$_7$ has gained attention as a quantum spin ice candidate with dipole-octupole character that arises from the crystal electric field ground state doublet for the Ce$^{3+}$ Kramers ion. This dipole-octupole character permits both spin-ice phases based on magnetic dipoles and those based on more-exotic octupoles. This work reports low-temperature neutron diffraction measurements on single crystal Ce$_2$Zr$_2$O$_7$ with $Q$-coverage both at low $Q$ where the magnetic form factor for dipoles is near maximal and at high $Q$ covering the region where the magnetic form factor for Ce$^{3+}$ octupoles is near maximal. This study was motivated by recent powder neutron diffraction studies of other Ce-based dipole-octupole pyrochlores, Ce$_2$Sn$_2$O$_7$ and Ce$_2$Hf$_2$O$_7$, which each showed temperature-dependent diffuse diffraction at high $Q$ that was interpreted as arising from octupolar correlations. Our measurements use an optimized single crystal diffuse scattering instrument that allows us to screen against strong Bragg scattering from Ce$_2$Zr$_2$O$_7$. The temperature-difference neutron diffraction reveals a low-$Q$ peak consistent with dipolar spin ice correlations reported in previous work, and an alternation between positive and negative net intensity at higher $Q$. These features are consistent with our numerical-linked-cluster calculations using pseudospin interaction parameters previously reported for Ce$_2$Zr$_2$O$_7$, Ce$_2$Sn$_2$O$_7$, and Ce$_2$Hf$_2$O$_7$. Importantly, neither the measured data nor any of the NLC calculations show evidence for increased scattering at high $Q$ resulting from octupolar correlations. We conclude that at the lowest attainable temperature for our measurements ($T = 0.05$~K), scattering from octupolar correlations in Ce$_2$Zr$_2$O$_7$ is not present in the neutron diffraction signal on the level of our observation threshold of $\sim$ 0.1$\%$ of the low-$Q$ dipole scattering. We compare these results to those obtained earlier on powder Ce$_2$Sn$_2$O$_7$ and Ce$_2$Hf$_2$O$_7$, and to low-energy inelastic neutron scattering from single crystal Ce$_2$Zr$_2$O$_7$.
\end{abstract}
\maketitle

\section{Introduction}
\label{Sec:I}
 
The study of magnetic materials with the potential for exotic ground states due to geometrical frustration is very topical in contemporary condensed matter physics due to the diversity and complexity of these ground states and the nature of their elementary excitations. Quantum spin liquid ground states are a class of quantum disordered ground states in which the magnetic spins or pseudospins remain disordered down to zero temperature and exhibit a high degree of quantum entanglement~\cite{Balents2010,Knolle2019}. Quantum spin ice (QSI) \cite{Hermele2004,GingrasReview2014} is a specific form of quantum spin liquid that exists on a pyrochlore lattice, a lattice of corner-sharing tetrahedra, for which the corresponding classical spin degeneracy at low temperature mimics that of proton disorder in water ice~\cite{Bramwell2001b,Castelnovo_12}. The theoretical study of QSIs has shown that such systems can be mapped onto the problem of quantum electrodynamics in three dimensions, with emergent low-energy elementary excitations being gapless photons, gapped electric monopoles (or spinons), and gapped magnetic monopoles (or visons)~\cite{Moe_trirvb_2003,Hermele2004, Benton2012, GingrasReview2014, Chen2023}. With this as motivation, much effort has been devoted to the experimental search for QSI phases in real materials, leading to numerous potential material realizations~\cite{Gardner1999, Zhou2008, Hiroshi2011, Kimura2013, Fritsch2014, Sibille2015, Sibille2016, Sarte2017, Sibille2018, Gao2018, Dalmau2019, Gaudet2019, Sibille2020, Smith2022, Changlani2022, Poree2022, Poree2023b, Tang_2023}. 

The insulating magnetic rare-earth cubic-pyrochlores are a family of materials in which the magnetic ions occupy a framework of corner-sharing tetrahedra. The $R_2B_2$O$_7$ cubic-pyrochlores constitute a large family of these magnetic insulators, where $R^{3+}$ is a magnetic rare-earth ion and $B^{4+}$ is a transition metal ion. Many can be synthesized as phase-pure powders or in single crystal form and these have been the subject of much experimental attention, leading to several tantalizing possible realizations of QSIs. These include studies of several non-Kramers pyrochlore systems such as Tb$_2$Ti$_2$O$_7$~\cite{Gardner1999, Hiroshi2011, Fritsch2014}, Pr$_2$Sn$_2$O$_7$~\cite{Zhou2008, Sarte2017}, Pr$_2$Zr$_2$O$_7$~\cite{Kimura2013,Tang_2023}, and Pr$_2$Hf$_2$O$_7$~\cite{Sibille2016, Sibille2018}. 

In particular, the cerium-based Kramers pyrochlores Ce$_2$Zr$_2$O$_7$, Ce$_2$Sn$_2$O$_7$, and Ce$_2$Hf$_2$O$_7$ have been intensively discussed as promising dipole-octupole QSI candidates~\cite{Sibille2015, Sibille2020, Gaudet2019, Gao2019, Smith2022, Changlani2022, Poree2022, Gao2022, Poree2023, Poree2023b, Smith2023, Beare2023, Yahne2024, Smith2025}. In each of the materials, the dipole-octupole symmetry of the Ce$^{3+}$ crystal electric field (CEF) ground state doublet permits a rich magnetic phase space containing numerous different types of quantum spin liquid and ordered ground states. In further detail, a description in terms of pseudospin is permitted by a CEF ground state that is well-separated in energy from the excited CEF states, and the symmetry of the CEF ground state dictates the general form of the pseudospin interaction Hamiltonian~\cite{Curnoe2007, Ross2011, Lee2012, Huang2014, RauReview2019}. For pyrochlores with a dipole-octupole CEF ground state doublet, the nearest-neighbor pseudospin-1/2 interaction Hamiltonian in zero external magnetic field is given by~\cite{Huang2014, RauReview2019}: 

\begin{equation}
\label{Eq:1}
\begin{split}
    \mathcal{H}_\mathrm{DO} & = \sum_{\langle ij \rangle}[     J_{x}{S_i}^{x}{S_j}^{x} + J_{y}{S_i}^{y}{S_j}^{y} + J_{z}{S_i}^{z}{S_j}^{z} \\ 
    & + J_{xz}({S_i}^{x}{S_j}^{z} + {S_i}^{z}{S_j}^{x})] \;,
\end{split}
\end{equation}

\noindent where ${S_{i}}^{\alpha}$ ($\alpha = x$, $y$, $z$) are the pseudospin components of the rare-earth atom $i$ in the local $\{x$, $y$, $z\}$ coordinate frame. This coordinate frame is defined locally for each ion $i$ with the $\hat{{\bf z}}_i$ anisotropy axis along the threefold rotation axis through rare-earth site $i$, with $\mathbf{y}_i$ along one of the symmetrically equivalent twofold rotation axes through rare-earth site $i$, and with $\hat{{\bf x}}_i = \hat{{\bf y}}_i \cross \hat{{\bf z}}_i$. 

The dipole-octupole symmetry of the CEF ground state doublet corresponds to the fact that $S^{x}$ and $S^{z}$ transform under time-reversal and site-symmetry operations like dipoles while $S^{y}$ transforms like the component of an octupole tensor~\cite{Huang2014,RauReview2019}. These symmetries are not to be confused with the physical magnetic moments carried by the pseudospin components, which are not necessarily the same as the symmetry: the lowest order multipolar moment carried by $S^{z}$ is a dipolar magnetic moment while for $S^{x}$ and $S^{y}$ it is an octupolar magnetic moment~\cite{Patri2020,Desrochers2022,Desrochers2024}.

This nearest-neighbor exchange Hamiltonian [\autoref{Eq:1}] can then be simplified via rotation of each local $\{x, y, z\}$ coordinate frame by $\theta$ about the respective local $y$-axis, where $\theta$ is given by $\theta = \frac{1}{2}\tan^{-1}[2J_{xz}/({J_{x}-J_{z}})]$~\cite{Huang2014, Benton2016}. These rotations yield new local coordinate frames, which are commonly denoted as the local $\{\tilde{x},\tilde{y},\tilde{z}\}$ coordinate frames, and the new nearest-neighbor Hamiltonian in these local coordinate frames is the ``XYZ'' Hamiltonian~\cite{Huang2014}:

\begin{equation} \label{Eq:2}
\begin{split}
    \mathcal{H}_\mathrm{XYZ} & = \sum_{\langle ij \rangle}[     J_{\tilde{x}}{S_i}^{\tilde{x}}{S_j}^{\tilde{x}} + J_{\tilde{y}}{S_i}^{\tilde{y}}{S_j}^{\tilde{y}} + J_{\tilde{z}}{S_i}^{\tilde{z}}{S_j}^{\tilde{z}}] \;.
\end{split}
\end{equation}

Theoretical work has shown that this zero-field XYZ Hamiltonian [\autoref{Eq:2}] has a rich magnetic ground state phase diagram containing all-in all-out dipole order and octupole order, as well as ``0-flux'' and ``$\pi$-flux'' QSIs with dipole-based and octupole-based versions of each~\cite{Huang2018b, Benton2020, Patri2020, Kim2022, Desrochers2022, Smith2023}. Furthermore, a quantum Monte Carlo study has shown evidence for a possible $\mathbb{Z}_2$ quantum spin liquid ground state along the border between the ordered and disordered regime of the zero-field ground state phase diagram~\cite{Huang2020}. In addition to this, several recent works using the nearest-neighbor Heisenberg model ($J_{\tilde{x}} = J_{\tilde{y}} = J_{\tilde{z}} = J$) for $J>0$ suggest the possibility of a novel symmetry-broken phase at that point in phase space~\cite{hagymasi_possibile_2021,astrakhantsev_broken-symmetry_2021,abundance_schaefer_2023,pohle_ground_2023}.

Recent powder neutron diffraction studies on samples of Ce$_2$Sn$_2$O$_7$ grown through solid-state synthesis techniques report temperature-dependent diffuse neutron diffraction at high $Q$, which was attributed to scattering from magnetic octupole moments. There, the authors further interpret their data in terms of an octupole-based $\pi$-flux quantum spin ice~\cite{Sibille2020}. However, a similar powder neutron diffraction study on hydrothermally-grown samples of Ce$_2$Sn$_2$O$_7$ in Ref.~\cite{Yahne2024} finds no such signal at high $Q$. Further, fits of the measured heat capacity, magnetic susceptibility and diffuse neutron scattering from this hydrothermally-grown Ce$_2$Sn$_2$O$_7$ sample to the relevant theory yielded estimates of the exchange parameters in the XYZ Hamiltonian relevant to dipole-octupole pyrochlores at the nearest neighbor level. This work placed the ground state of this Ce$_2$Sn$_2$O$_7$ sample as a dipole-based all-in, all-out magnetic phase that is proximate to a dipole-based 0-flux quantum spin ice phase. Interestingly, both Ref.~\cite{Sibille2020} and Ref.~\cite{Yahne2024} find a low-$Q$ scattering signal typical of spin ice correlations between dipole moments. Additionally, recent measurements on Ce$_2$Hf$_2$O$_7$ have been interpreted in terms of a $\pi$-flux quantum spin ice~\cite{Poree2022, Poree2023b}, with a collection of experimental data fit to constrain the nearest-neighbor exchange parameters and concluding a dominant $J_{\tilde{x}}$ or $J_{\tilde{y}}$ in the XYZ Hamiltonian for Ce$_2$Hf$_2$O$_7$.

In this work, we focus on the QSI candidate Ce$_2$Zr$_2$O$_7$, where the collection of existing measurements is consistent with a lack of both magnetic order and spin freezing above $T \sim 0.02$~K in zero external magnetic field~\cite{Gaudet2019, Gao2019, Gao2022, Smith2022, Beare2023}. Fits of experimental data to theory allow estimates of the exchange parameters in the XYZ Hamiltonian for Ce$_2$Zr$_2$O$_7$, with the resulting parameters suggesting a $\pi$-flux QSI ground state~\cite{Smith2022, Changlani2022, Smith2023}. Refs.~\cite{Smith2022, Smith2023} perform fits to collections of experimental data using numerical-linked-cluster (NLC) calculations~\cite{tang_nlce_2013,Schafer2020,schaefer_magnetic_2022}, with each of these works indicating a $\pi$-flux quantum spin ice ground state on the boundary between dipolar and octupolar character. However, Ref.~\cite{Changlani2022} fits a separate collection of experimental data on a different single crystal using the finite-temperature Lanczos method and this work finds exchange parameters corresponding to an octupole-based $\pi$-flux quantum spin ice ground state. Importantly, the resulting exchange parameters from these different fittings for Ce$_2$Zr$_2$O$_7$ are relatively near to each other in parameter space despite the slight difference in the corresponding magnetic ground states suggested by Refs.~\cite{Smith2022, Smith2023} and Ref.~\cite{Changlani2022}.

The large majority of measurements previously published on Ce$_2$Zr$_2$O$_7$ have little direct sensitivity to the magnetic octupole moments or their correlations, with heat capacity measurements being a notable exception~\cite{Gao2019, Smith2022, Gao2022, Smith2023, Chen2023}. The heat capacity measurements on Ce$_2$Zr$_2$O$_7$ show no signs of magnetic order for either the dipoles or octupoles. The subsequent fitting of the exchange parameters in the XYZ Hamiltonian to these measurements, along with fits to the magnetic susceptibility and magnetization, has allowed for theory to infer the behavior of both the dipole and octupole moments in Ce$_2$Zr$_2$O$_7$. Nonetheless, it remains very desirable to expand available measurements on Ce$_2$Zr$_2$O$_7$ such that more information with direct sensitivity to the octupolar moments and their correlations is brought to bear in this field, and that is the main focus of this work. 

Specifically, this work uses forefront neutron scattering techniques on single crystal Ce$_2$Zr$_2$O$_7$ in search for magnetic octupolar correlations. Here by octupolar correlations we mean two-point correlation functions of on-site magnetic octupole moments, or two-point correlations involving $S^{x}$ components or $S^{y}$ components in terms of pseudospin, as opposed to higher order correlations of on-site dipoles~\cite{Zhitomirsky2008}. In further detail, we report neutron scattering measurements with relatively high incident energy and a correspondingly large $Q$-range, allowing for $Q$-coverage well past $Q \sim 8$~$\angstrom^{-1}$, where the magnetic form factor for Ce$^{3+}$ octupoles is near maximal~\cite{PlackeThesis}. While similar to the earlier high-energy neutron scattering measurements on powder samples of Ce$_2$Sn$_2$O$_7$ and Ce$_2$Hf$_2$O$_7$~\cite{Sibille2020, Poree2023b}, with a large $Q$-range, the present neutron scattering measurements were performed on a single crystal sample of Ce$_2$Zr$_2$O$_7$, using the dedicated diffuse scattering spectrometer CORELLI at the Spallation Neutron Source. This provides vector-{\bf Q} scattering information in a three-dimensional volume of reciprocal space, which allows for great flexibility in, for example, avoiding the very strong single crystal Bragg scattering from the Ce$_2$Zr$_2$O$_7$ sample.

\section{Experimental Details}
\label{Sec:II}

Neutron scattering measurements were performed on a high-quality 1.5~gram single crystal sample of Ce$_2$Zr$_2$O$_7$ grown via floating zone image furnace techniques as described in Ref.~\cite{Gaudet2019}. As also described in Refs.~\cite{Gaudet2019, Smith2022, Smith2023, Beare2023}, sample oxidation and accompanying non-magnetic Ce$^{4+}$ impurities can complicate the interpretation of measurements performed on as-grown Ce$_2$Zr$_2$O$_7$ samples. Furthermore, significant sample oxidation occurs for Ce$_2$Zr$_2$O$_7$ left in air at room temperature~\cite{Gaudet2019}, at least for powder samples. Accordingly, and similar to the process described in Refs.~\cite{Gaudet2019, Smith2022, Smith2023, Beare2023}, our crystal was annealed at 1400~$^\circ$C for 72~hours in H$_2$ gas to reduce the as-grown oxygen and Ce$^{4+}$ content. Samples were stored in inert gas after annealing until being loaded into the inert environment used for measurement. 

Our neutron scattering measurements employed the CORELLI diffuse scattering diffractometer at the Spallation Neutron Source of Oak Ridge National Laboratory, which employs neutrons over the range of incident energies from $E_i = 10$~meV to $E_i = 200$~meV and uses a cross-correlation chopper to give an energy discrimination centered on zero energy transfer with energy resolution that is roughly 3-5\% of the incident neutron energy $E_i$. The energy resolution reaches 2.5 meV at $E_i$ = 50~meV, which is at the peak of the flux spectrum. We henceforth refer to these neutron scattering measurements, with variable energy resolution, as neutron diffraction measurements, as this relatively wide energy resolution is expected to capture any low-lying dynamic magnetic signal from Ce$_2$Zr$_2$O$_7$ along with the elastic signal~\cite{Gaudet2019, Smith2022, Gao2022, Smith2023}. Specifically, the dynamic neutron scattering signal at energy transfer around $E = 0.1$~meV, attributed to dipolar spin ice correlations in Refs.~\cite{Gaudet2019, Smith2022, Smith2023}, is well captured by the energy resolution. The 1.5~gram single crystal sample of Ce$_2$Zr$_2$O$_7$ was mounted in a copper sample holder and aligned with the $(H, H, L)$ plane coincident with the horizontal plane, as to give sample rotation about the vertical $(K, \bar{K}, 0)$ axis. The resulting measurements cover a volume of reciprocal space centered on the $(H, H, L)$ plane. A dilution refrigerator was used to reach temperatures as low as $T = 0.05$~K. For each chosen temperature of our neutron scattering experiment, the sample was rotated in the $(H,H,L)$ plane in 1$^\circ$ steps through a total of 360$^\circ$. 

Nuclear Bragg peaks from the Ce$_2$Zr$_2$O$_7$ sample are relatively strong and possess a weak temperature-dependence, making the regions in which they appear less useful for identifying temperature-dependent, low-intensity diffuse scattering due to magnetic correlations. Due to this, we focus on a region of reciprocal space that does not contain nuclear Bragg peaks from the Ce$_2$Zr$_2$O$_7$ sample. Specifically, the vertical span of CORELLI's detector allows for examination of the signal measured from Ce$_2$Zr$_2$O$_7$ in the $(H+0.5,H-0.5,L)$ plane of reciprocal space, which is shifted away from the $(H,H,L)$ plane as to avoid the nuclear Bragg peaks from Ce$_2$Zr$_2$O$_7$, which occur for all even or all odd $h,k,l$ indices in $\mathbf{Q} = (h,k,l)$. This is particularly important for measurements covering high-$Q$ scattering because the number of nuclear Bragg peaks encountered grows rapidly with $Q$. 

The nuclear Bragg peaks are also the zone centers for acoustic phonons, which definitely contribute an unwanted thermal component to the diffuse scattering. The present measurements using CORELLI's correlation chopper greatly reduce this phonon scattering compared with corresponding measurements using a powder neutron diffractometer (as was employed in earlier studies~\cite{Sibille2020, Poree2023b}). However, this potential background contribution to the diffuse scattering of interest is also minimized by our use of the $(H+0.5,H-0.5,L)$ plane and the corresponding exclusion of nuclear Bragg peaks from single crystal Ce$_2$Zr$_2$O$_7$. 

The large volume of reciprocal space from CORELLI will also capture powder scattering from the Al and Cu addenda associated with the cryostat and sample mount. This source of background scattering is less intense than the single crystal Bragg intensity from the sample, but it remains much stronger than the diffuse scattering of interest. Further, as this powder scattering depends only on $Q = \|\mathbf{Q}\|$, it cannot be avoided by choosing low symmetry $\mathbf{Q}$ integrations of the CORELLI datasets. This is similarly unavoidable in powder neutron diffraction measurements. The effect of this background scattering would, in principle, be reduced with a subtraction of a high-temperature dataset, as will be employed in this study. However, this strong scattering has a weak temperature dependence to it, and the imperfect subtraction of these strong signals corrupts the identification of temperature-dependent diffuse scattering of interest. For this reason, we identify and mask out each neutron powder ring in the dataset and analyze only the data that is not corrupted by the subtraction of these large signals. These regions can be easily identified as white rings in Figs.~\ref{Figure2}, \ref{Figure3} and \ref{Figure4}. 

To reduce the effects of any anisotropic background scattering and neutron absorption, we utilize a symmetrization of the diffraction data that takes advantage of the mirror-plane symmetries relevant to the cubic reciprocal lattice of the pyrochlores. This symmetrization enforces mirror-plane symmetries on the measured data in order to wash out anisotropic background scattering and neutron absorption effects (which do not adhere to these symmetries) compared to scattering from the single crystal sample (which does adhere to these symmetries). Specifically, this symmetrization is used in Figs.~\ref{Figure2}, \ref{Figure3}(d-f), and \ref{Figure4}(d-f). This symmetrization process is further discussed in the supplemental material of Ref.~\cite{Gaudet2019}.

The Data Analysis and Visualization Environment (DAVE) software suite for the reduction, visualization, and analysis of low-energy neutron spectroscopic data (Ref.~\cite{Azuah2009}) was used in analyzing the neutron scattering data presented in this paper.

\section{Results}
\label{Sec:III}

%%%%%%%%%%%%%%%%%%%%%%%%%%%%%%%%%

\begin{figure}[t]
\linespread{1}
\par
\includegraphics[width=3.4in]{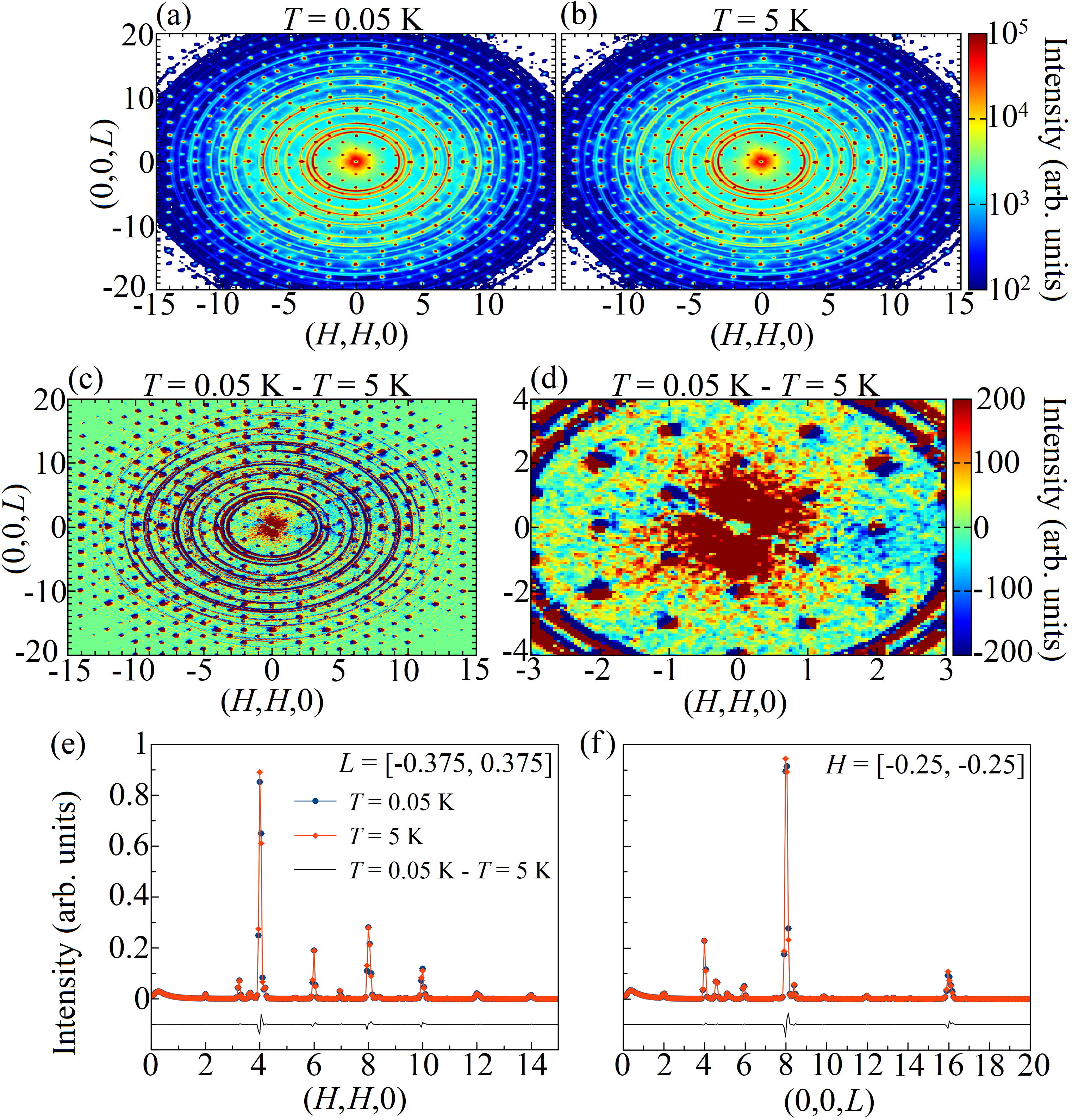}
\par
\caption{The neutron diffraction signal measured from Ce$_2$Zr$_2$O$_7$ in the $(H,H,L)$ plane of reciprocal space for an integration in $(K,\bar{K},0)$ over $K = [-0.2, 0.2]$. Specifically, we show the measured neutron diffraction in the $(H,H,L)$ plane at (a)~$T = 0.05$~K, (b)~$T = 5$~K, and (c,d)~for $T = 0.05$~K with the $T = 5$~K data subtracted. (e)~The neutron diffraction measured from Ce$_2$Zr$_2$O$_7$ along the $(H,H,0)$ direction for an integration in $(0,0,L)$ over $L = [-0.375, 0.375]$. (f)~The neutron diffraction measured from Ce$_2$Zr$_2$O$_7$ along the $(0,0,L)$ direction for an integration in $(H,H,0)$ over $H = [-0.25, 0.25]$. The $T = 0.05$~K$~-~T = 5$~K temperature-difference data has been shifted downwards by 0.1 units for visibility in both (e) and (f).} 
\label{Figure1}
\end{figure}

%%%%%%%%%%%%%%%%%%%%%%%%%%%%%%%%%

Figure~\ref{Figure1} shows the neutron diffraction pattern measured from Ce$_2$Zr$_2$O$_7$ in the $(H,H,L)$ plane of reciprocal space, with a symmetric integration in the third dimension $(K,\bar{K},0)$ over $K = [-0.2, 0.2]$ at both $T = 0.05$~K [Fig.~\ref{Figure1}(a)] and $T = 5$~K [Fig.~\ref{Figure1}(b)]. The intensity-scale is logarithmic in each case. As shown in Figs.~\ref{Figure1}(a,b), this $(H,H,L)$ plane of reciprocal space contains a multitude of nuclear Bragg peaks from single crystal Ce$_2$Zr$_2$O$_7$, nuclear Bragg powder rings from scattering originating from the sample holder and sample environment equipment, and weak diffuse scattering which is much weaker in intensity than the aforementioned Bragg contributions. To help isolate any magnetic dipolar or octupolar contribution to the measured diffuse scattering at $T = 0.05$~K, we subtract the data measured at $T = 5$~K and show the $T = 0.05$~K$~-~T = 5$~K temperature subtraction in Fig.~\ref{Figure1}(c,d) with Fig.~\ref{Figure1}(c) showing a large region of reciprocal space and with Fig.~\ref{Figure1}(d) focusing on the low-$Q$ region where diffuse scattering from magnetic dipoles is expected to strongly dominate any scattering from magnetic octupoles due to the difference in size of the corresponding magnetic form factors for dipoles and octupoles~\cite{Smith2022Reply, PlackeThesis}. In Fig.~\ref{Figure1}(c,d), this net intensity is on a linear scale and is symmetric around zero with a much smaller range than that employed for the unsubtracted datasets in Fig.~\ref{Figure1}(a,b). While it is clear that diffuse net intensity due to dipolar correlations is present at small $Q$, even on this lower intensity-scale, it is not immediately obvious whether significant diffuse scattering is present at high $Q$, due to the large number of nuclear Bragg peaks that interfere with the visual search for the octupolar diffuse scattering at these high wavevectors. 

The Bragg peak locations in the $T = 0.05$~K$~-~T = 5$~K temperature subtraction of Fig.~\ref{Figure1}(c,d) each shows a combination of positive and negative intensity, resulting from the imperfect subtraction of the high-intensity nuclear Bragg peaks at those locations. This is consistent with both the lack of magnetic Bragg scattering reported for Ce$_2$Zr$_2$O$_7$ in zero field at low temperature in Refs.~\cite{Gao2019, Gaudet2019, Smith2022, Gao2022, Smith2023} and with expectations for a material that has a quantum spin ice ground state. Nonetheless, we take a closer look for any possible magnetic Bragg scattering from Ce$_2$Zr$_2$O$_7$ at $T = 0.05$~K using the $\mathbf{Q}$-dependence of the neutron diffraction data from Ce$_2$Zr$_2$O$_7$ along high-symmetry directions of reciprocal space. Figs.~\ref{Figure1}(e) and 1(f) show the $\mathbf{Q}$-dependence at both $T = 0.05$~K and $T = 5$~K for $\mathbf{Q}$ along the $(H,H,0)$ and $(0,0,L)$ directions of reciprocal space, respectively. As shown in Figs.~\ref{Figure1}(e,f), we observe no clear increase in intensity for any Bragg peak locations at $T = 0.05$~K (blue) compared to $T = 5$~K (red). This is shown by the $T = 0.05$~K$~-~T = 5$~K temperature subtraction (black), where we observe only minor differences in the intensity values at Bragg peak locations, which we attribute to statistical counting error and error associated with small shifts of the sample with changing temperature.  

Due to the large number of nuclear Bragg peaks from Ce$_2$Zr$_2$O$_7$ in the $(H,H,L)$ plane, we shift our search for diffuse scattering to the $(H+0.5,H-0.5,L)$ plane within our three-dimensional volume of measured reciprocal space. This plane is parallel to the $(H,H,L)$ but shifted along the orthogonal $(K,\bar{K}, 0)$ direction by $K = 0.5$ so as to avoid nuclear Bragg peaks from single crystal Ce$_2$Zr$_2$O$_7$. As discussed above, this shift in the $(K,\bar{K}, 0)$ direction also helps to avoid scattering from low-energy acoustic phonons, which emanate from Bragg peak locations. Specifically, we use an integration in $(K,\bar{K}, 0)$ over the range $K = [-0.7, -0.3]$ and the symmetrically-equivalent range $K = [0.3, 0.7]$, and we henceforth denote this as the $(H+0.5,H-0.5,L)$ plane.

%%%%%%%%%%%%%%%%%%%%%%%%%%%%%%%%%

\begin{figure}[]
\linespread{1}
\par
\includegraphics[width=3.4in]{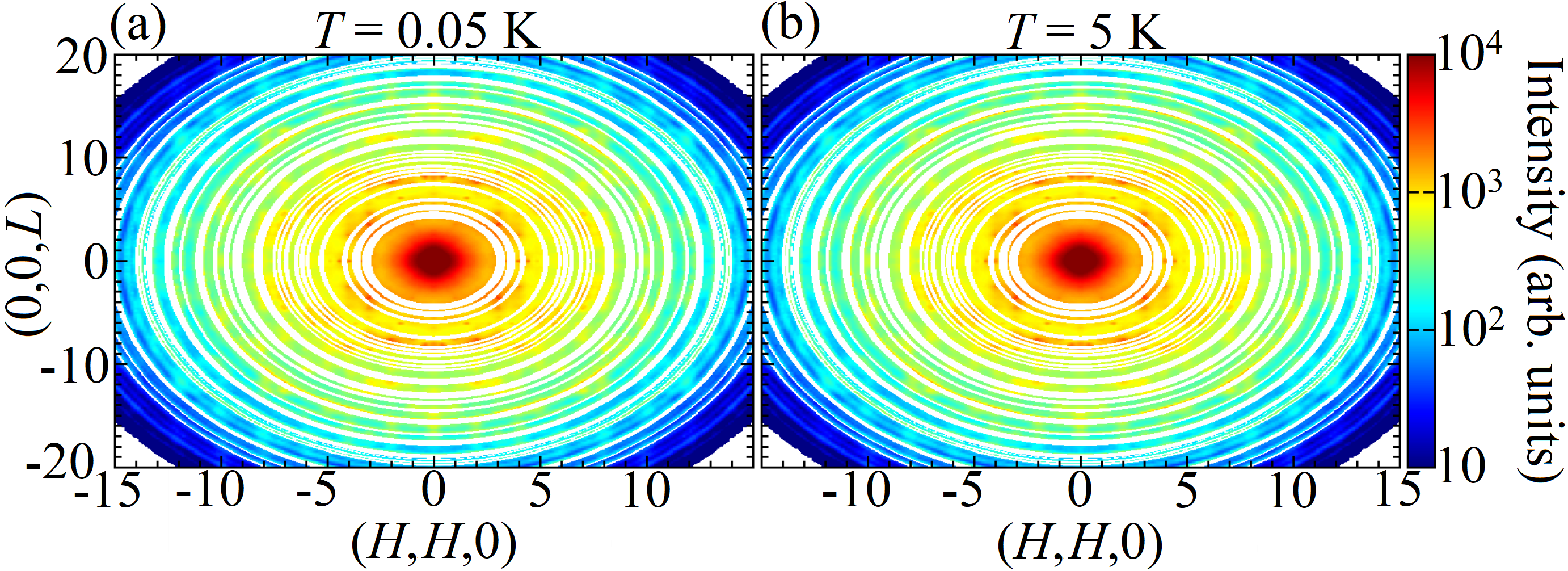}
\par
\caption{The symmetrized neutron diffraction signal measured from Ce$_2$Zr$_2$O$_7$ in the $(H,H,L)$ plane of reciprocal space for an integration in $(K,\bar{K},0)$ over $K = [-0.7, -0.3]$ and $K = [0.3, 0.7]$, which we refer to as the $(H+0.5,H-0.5,L)$ plane. Specifically, we show the measured neutron diffraction in the $(H+0.5,H-0.5,L)$ plane at (a)~$T = 0.05$~K and (b)~$T = 5$~K with the powder rings from the sample holder and sample environment equipment removed from the data. } 
\label{Figure2}
\end{figure}
%%%%%%%%%%%%%%%%%%%%%%%%%%%%%%%%%
%%%%%%%%%%%%%%%%%%%%%%%%%%%%%%%%%

\begin{figure*}[]
\linespread{1}
\par
\includegraphics[width=5.5in]{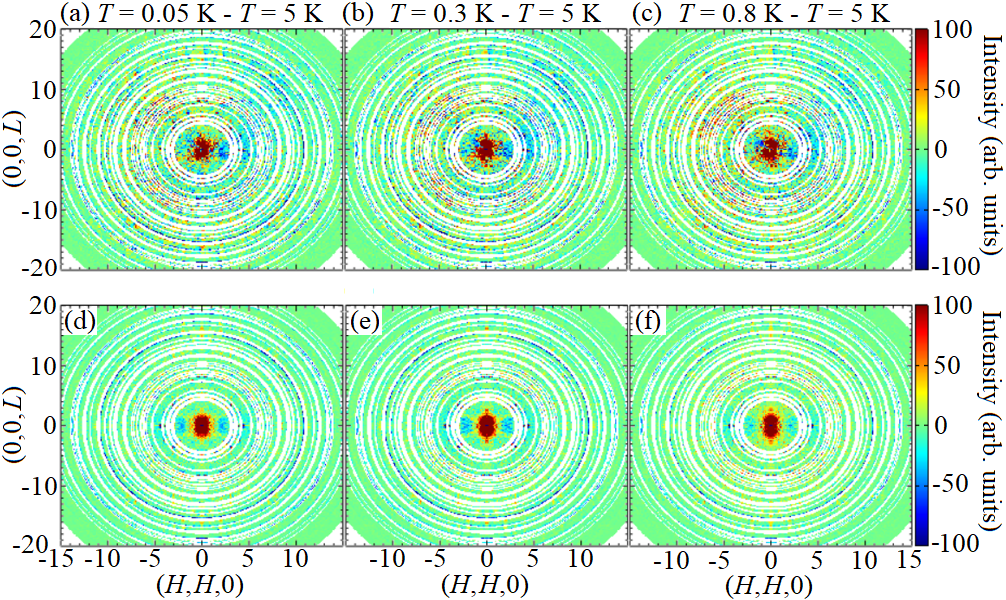}
\par
\caption{The neutron diffraction signal measured from Ce$_2$Zr$_2$O$_7$ in the $(H,H,L)$ plane of reciprocal space for an integration in $(K,\bar{K},0)$ over $K = [-0.7, -0.3]$ and $K = [0.3, 0.7]$, which we refer to as the $(H+0.5,H-0.5,L)$ plane. Specifically, we show the measured temperature-difference neutron diffraction in the $(H+0.5,H-0.5,L)$ plane for (a,d)~$T = 0.05$~K, (b,d)~$T = 0.3$~K, and (c,f)~$T = 0.8$~K with a $T = 5$~K dataset subtracted in each case and with the powder rings from the sample holder and sample environment equipment removed from the data. The data in (a-c) contains no additional symmetrization, while the data in (d-f) has been symmetrized (see main text).} 
\label{Figure3}
\end{figure*}

%%%%%%%%%%%%%%%%%%%%%%%%%%%%%%%%%

This new three-dimensional dataset in the $(H+0.5,H-0.5,L)$ plane does not contain any nuclear Bragg peaks from the single crystal Ce$_2$Zr$_2$O$_7$ sample. However, it does contain intensity from neutron powder diffraction rings from the sample holder and sample environment, and we identify and mask this scattering from the datasets so that we can examine the diffuse scattering without any Bragg scattering present. This is shown in Figure~\ref{Figure2} where we show similar patterns of diffuse scattering measured in the $(H+0.5,H-0.5,L)$ plane from Ce$_2$Zr$_2$O$_7$ at both $T = 0.05$~K [Fig.~\ref{Figure2}(a)] and $T = 5$~K [Fig.~\ref{Figure2}(b)]. Note that the intensity-scale is logarithmic and goes from $10^1$ to $10^4$ in Fig.~\ref{Figure2}, while the intensity-scale of the corresponding diffraction data in Figs.~\ref{Figure1}(a) and (b) went from $10^2$ to $10^5$. For that reason, it is possible to observe more diffuse scattering in Fig.~\ref{Figure2}. However, much of this diffuse scattering is clearly temperature-independent between $T = 0.05$~K and $T = 5$~K, from temperatures well below the estimated strength of magnetic interactions in Ce$_2$Zr$_2$O$_7$ to temperatures well above it~\cite{Changlani2022, Smith2022, Smith2023, Beare2023}. We attribute this temperature-independent diffuse scattering in Figure~\ref{Figure2} to nuclear diffuse scattering, likely related to the known oxygen stoichiometry issues in Ce$_2$Zr$_2$O$_7$, which are reduced by our hydrogen annealing protocol (see Section~\ref{Sec:II}) but remain present at a relatively minor level~\cite{Gaudet2019, Smith2022, Beare2023}.

In order to identify temperature-dependent diffuse neutron diffraction from magnetic correlations, we examine the difference between diffuse scattering at low temperature ($T = 0.05$~K, $T = 0.3$~K, and $T = 0.8$~K) and high temperature ($T= 5$~K). These subtracted diffuse scattering datasets in the $(H+0.5,H-0.5,L)$ plane  are shown in Fig.~\ref{Figure3}, both with no additional symmetrization of the data [Fig.~\ref{Figure3}(a-c)] and with mirror plane symmetry operations (see Experimental Details) applied to the data [Fig.~\ref{Figure3}(d-f)]. 

%%%%%%%%%%%%%%%%%%%%%%%%%%%%%%%%%

\begin{figure*}[t]
\linespread{1}
\par
\includegraphics[width=5.5in]{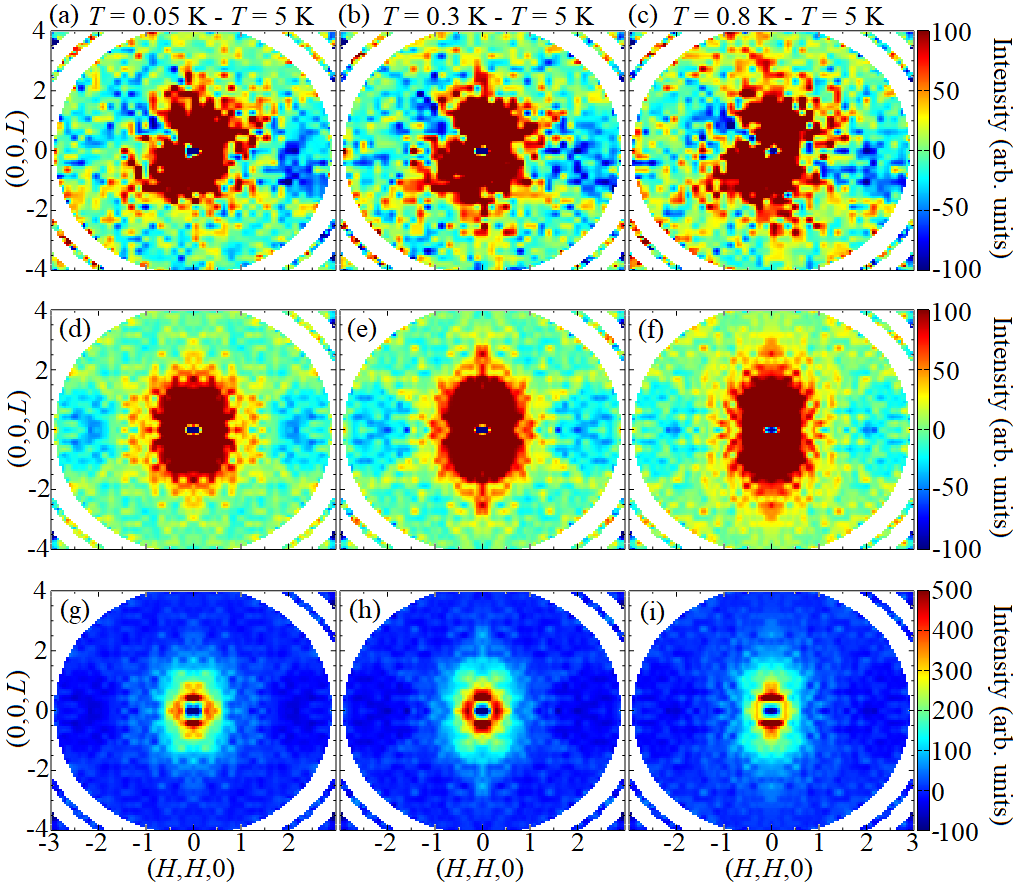}
\par
\caption{The low-$Q$ neutron diffraction signal measured from Ce$_2$Zr$_2$O$_7$ in the $(H,H,L)$ plane of reciprocal space for an integration in $(K,\bar{K},0)$ over $K = [-0.7, -0.3]$ and $K = [0.3, 0.7]$, which we refer to as the $(H+0.5,H-0.5,L)$ plane. Specifically, we show the measured temperature-difference neutron diffraction in the $(H+0.5,H-0.5,L)$ plane for (a,d,g)~$T = 0.05$~K, (b,e,h)~$T = 0.3$~K, and (c,f,i)~$T = 0.8$~K with a $T = 5$~K dataset subtracted in each case and with the powder rings from the sample holder and sample environment equipment removed from the data. The data in (a-c) contains no additional symmetrization while the data in (d-i) has been symmetrized (see main text). The intensity color scale for (a-f) ranges from -100 to 100, and for (g-i), we use an intensity-scale with a higher maximum, ranging from -100 to 500, to show the scattering near $Q = 0$, which is saturated on the intensity-scale of (a-f). Note that these datasets are shown with narrower $(H,H,0)$ and $(0,0,L$) ranges than in Fig.~\ref{Figure3}.} 
\label{Figure4}
\end{figure*}

%%%%%%%%%%%%%%%%%%%%%%%%%%%%%%%%%

As can be seen in Figure~\ref{Figure3}, diffuse scattering with relatively high net intensity is detected in the low-$Q$ region of reciprocal space. This is examined in greater detail in Fig.~\ref{Figure4} where a close-up of this low-$Q$ region of the temperature-difference data is shown. Figure~\ref{Figure4}(a-c) again show the temperature-difference data without additional symmetrization, while Fig.~\ref{Figure4}(d-f) show the corresponding data with symmetrization. We also show this symmetrized data for the three temperature subtractions in Fig.~\ref{Figure4}(g-i) with an increased maximum of the intensity color scale as to provide a clear view of the scattering near $Q=0$ that is saturated on the narrower intensity-scale of Fig.~\ref{Figure4}(a-f). The symmetrized diffuse scattering datasets at the lowest temperatures in Fig.~\ref{Figure4}(d,e,g,h) show positive diffuse net scattering extending out along $(0,0,1)$, $(1,1,1)$, and equivalent directions, strongly resembling the ``snowflake" pattern previously observed at low ${Q}$ from Ce$_2$Zr$_2$O$_7$ in Refs.~\cite{Gaudet2019, Gao2019, Smith2022}. This includes the broad peaks in the diffuse scattering centered on $H = 0$ near $Q \sim 0.55~\angstrom^{-1}$, which is near $L = 0.5$ in the $(H+0.5, H-0.5, L)$ plane [Figs.~\ref{Figure4}(g,h)] and near $L = 1$ in the $(H,H,L)$ plane examined in Refs.~\cite{Gaudet2019, Gao2019, Smith2022}, in each case with broader width along $(H,H,0)$ than along $(0,0,L)$. We therefore identify it as originating from the magnetic dipolar correlations appropriate to a material with a $\pi$-flux quantum spin ice ground state~\cite{Smith2022, Smith2023, Changlani2022}, with the estimated exchange parameters for Ce$_2$Zr$_2$O$_7$ in Refs.~\cite{Smith2022, Smith2023, Changlani2022} suggesting that spinons provide the dominant contribution to this low-$Q$ magnetic diffuse scattering.

%%%%%%%%%%%%%%%%%%%%%%%%%%%%%%%%%

\begin{figure*}[t]
\linespread{1}
\par
\includegraphics[width=6in]{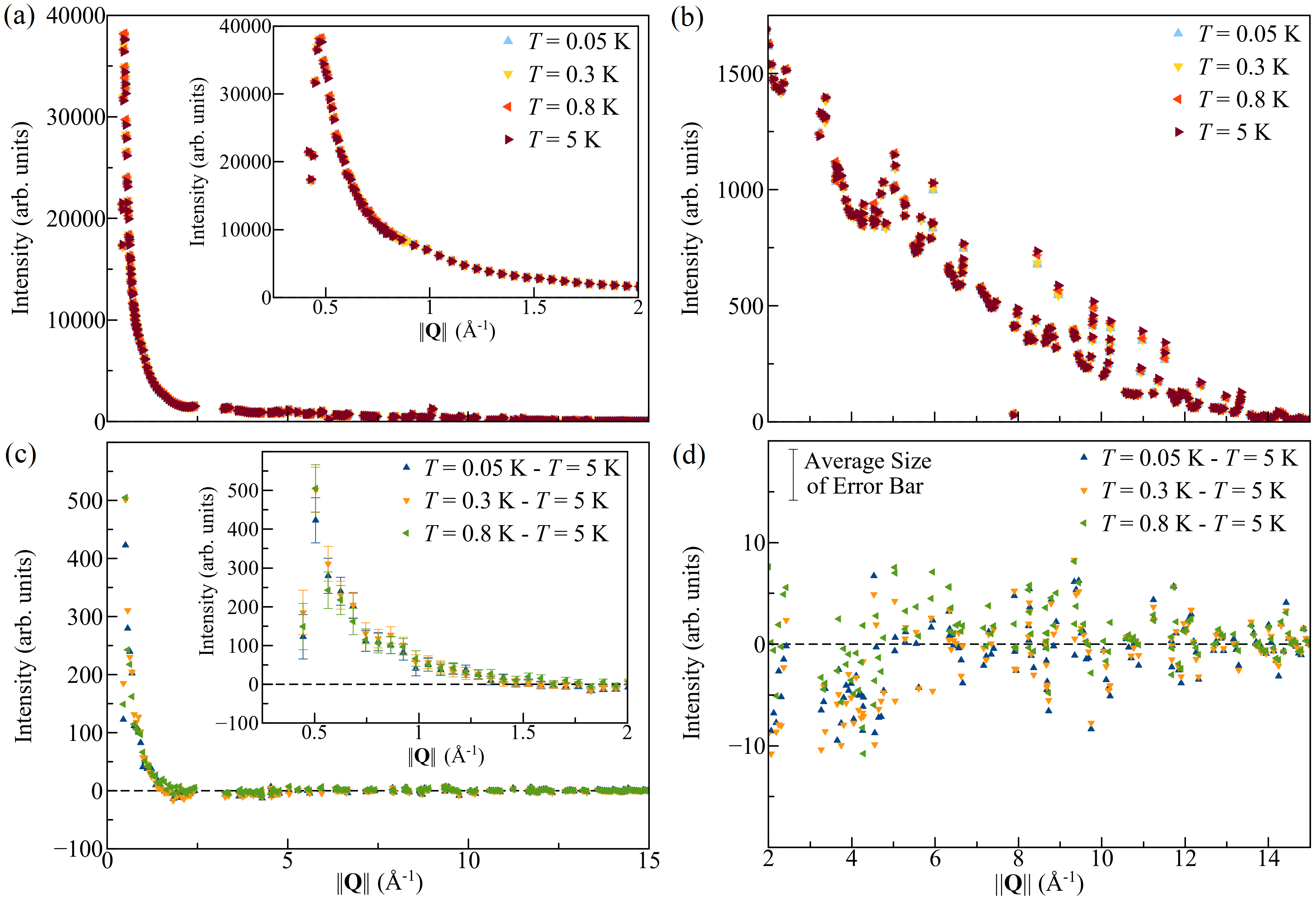}
\par
\caption{The powder-averaged neutron diffraction signal measured from single crystal Ce$_2$Zr$_2$O$_7$ in the $(H,H,L)$ plane of reciprocal space for an integration in $(K,\bar{K},0)$ over $K = [-0.7, -0.3]$ and $K = [0.3, 0.7]$, which we refer to as the $(H+0.5,H-0.5,L)$ plane. Specifically, we show this powder-average over the $(H+0.5,H-0.5,L)$ plane at $T = 0.05$~K, $T = 0.3$~K, $T = 0.8$~K, and $T = 5$~K in (a) and (b), and for the temperature-difference data $T = 0.05$~K~-~$T = 5$~K, $T = 0.3$~K~-~$T = 5$~K, and $T = 0.8$~K~-~$T = 5$~K in (c) and (d). The data points at locations of Bragg powder rings, associated with Bragg scattering from the sample holder and sample environment equipment, have been removed to ease the visual search for diffuse scattering between these locations.} 
\label{Figure5}
\end{figure*}

%%%%%%%%%%%%%%%%%%%%%%%%%%%%%%%%%

%%%%%%%%%%%%%%%%%%%%%%%%%%%%%%%%%
\begin{figure}[!h]
\linespread{1}
\par
\includegraphics[width=2.9in]{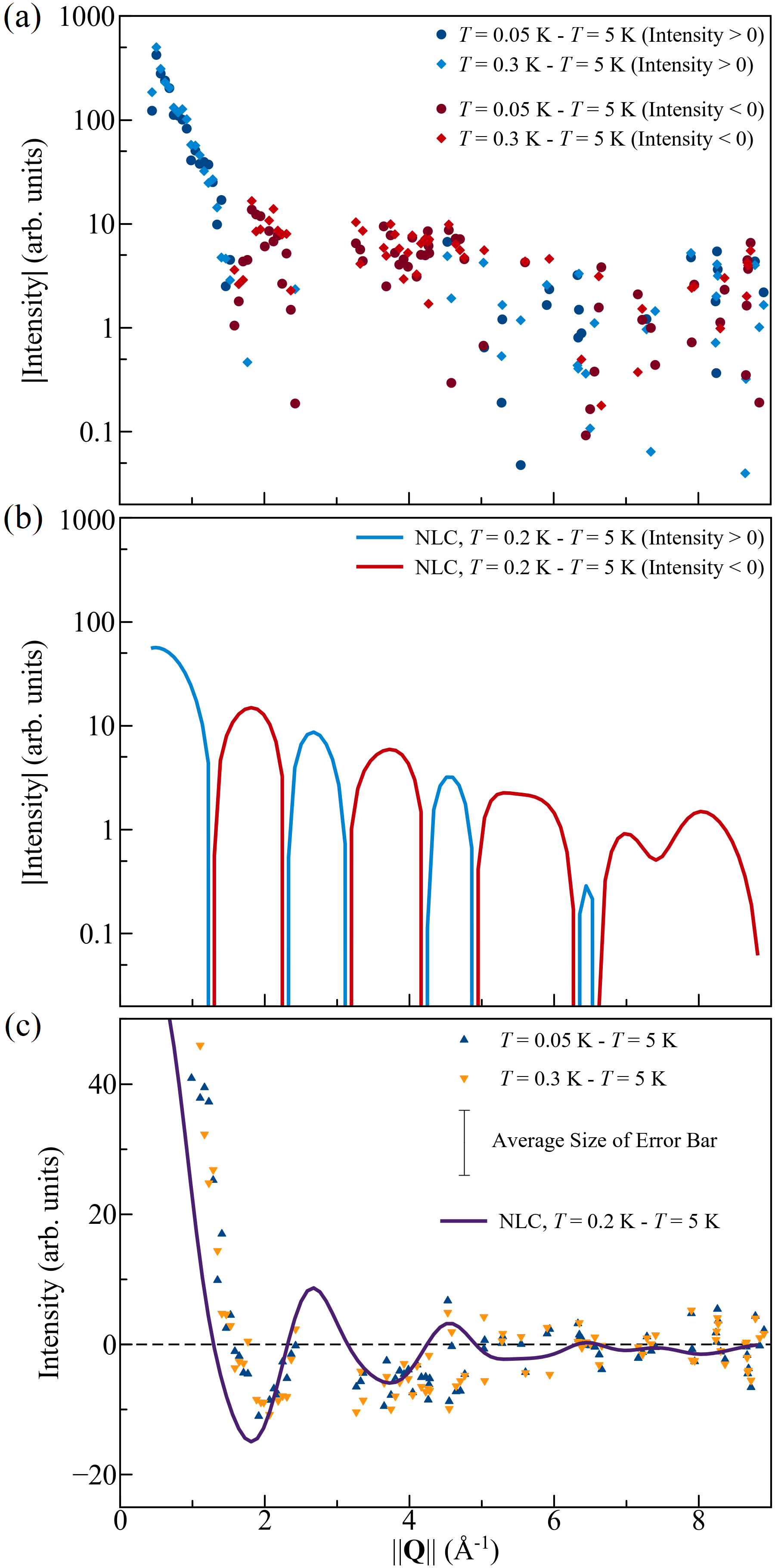}
\par
\caption{(a) The powder-average of temperature-difference neutron diffraction signal measured from single crystal Ce$_2$Zr$_2$O$_7$ in the $(H,H,L)$ plane of reciprocal space for an integration in $(K,\bar{K},0)$ over $K = [-0.7, -0.3]$ and $K = [0.3, 0.7]$, also shown in Fig.~\ref{Figure5}(c,d). Specifically, we show this powder-averaged data for $T = 0.05$~K (circles) and $T = 0.3$~K (diamonds) with subtraction of $T = 5$~K dataset in each case, on a logarithmic scale where points with positive net intensity are shown as blue and points with negative net intensity are shown as red. (b) Seventh-order NLC calculations for the powder averaged neutron diffraction intensity, with the powder average taken over the $(H,H,L)$ plane for an integration in $(K,\bar{K},0)$ over $K = [-0.7, -0.3]$ and $K = [0.3, 0.7]$, for $T = 0.2$~K with the corresponding $T = 5$~K calculation subtracted. Again, we use a logarithmic intensity-scale with positive (negative) net intensity shown as blue (red). (c) The same neutron scattering data and calculations shown in (a) and (b), respectively, on a linear intensity-scale. These NLC calculations employ the pseudospin Hamiltonian parameters estimated for Ce$_2$Zr$_2$O$_7$ in Ref.~\cite{Smith2022}.} 
\label{Figure6}
\end{figure}
%%%%%%%%%%%%%%%%%%%%%%%%%%%%%%%%%

%%%%%%%%%%%%%%%%%%%%%%%%%%%%%%%%%
\begin{figure}[!h]
    \centering
    \includegraphics[width=3.1in]{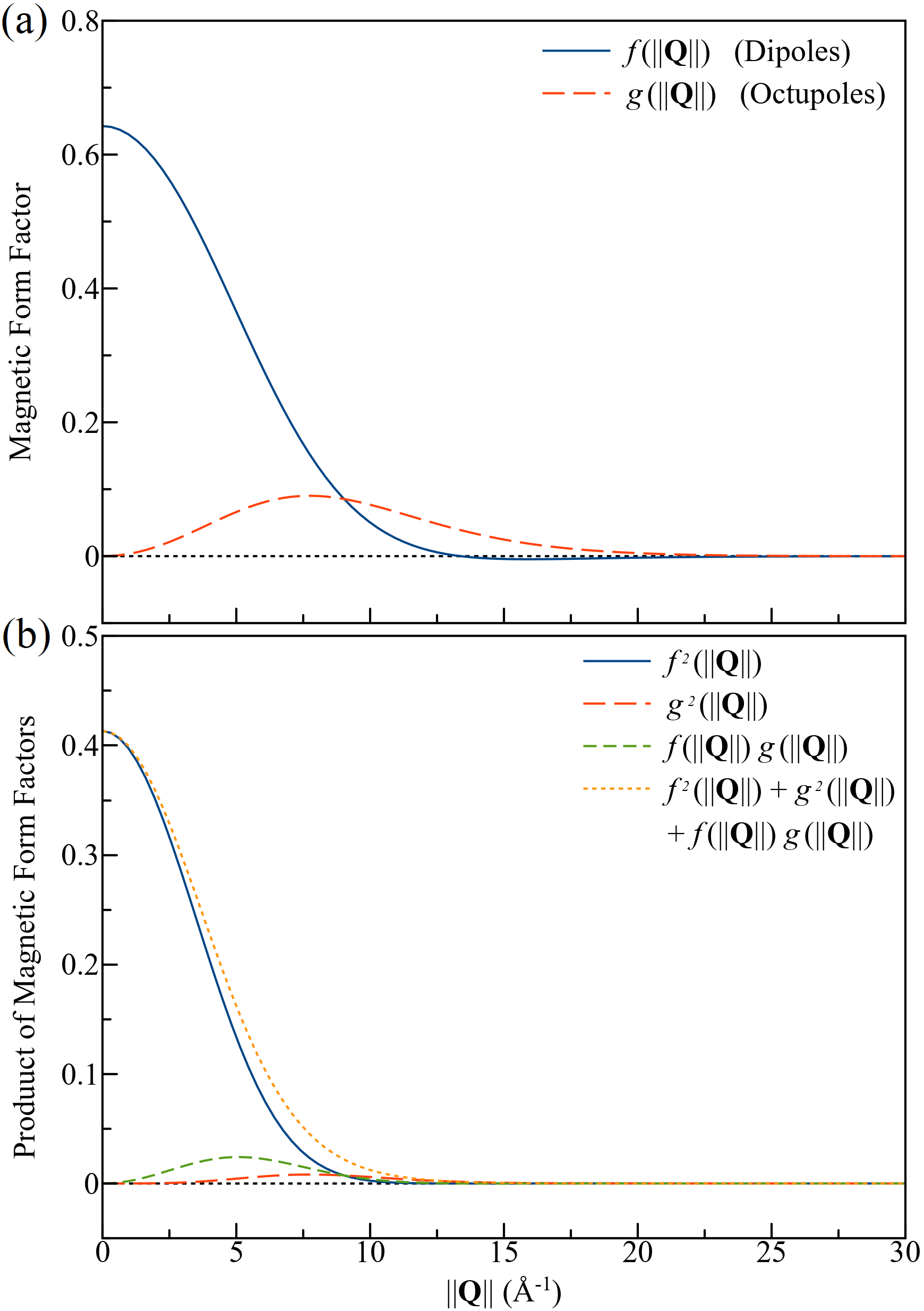}
    \caption{(a) The magnetic form factors for magnetic dipoles [($f(||\mathbf{Q}||)$] and octupoles [($g(||\mathbf{Q}||)$] in Ce$^{3+}$-based pyrochlores, calculated in Ref.~\cite{PlackeThesis}. (b) The relevant products of the magnetic form factors for magnetic dipoles and octupoles in Ce$^{3+}$-based pyrochlores~\cite{PlackeThesis}, as well as the sum of these products. Specifically, we show $f^2(||\mathbf{Q}||)$, $g^2(||\mathbf{Q}||)$, $f(||\mathbf{Q}||)g(||\mathbf{Q}||)$ and the sum of these. The sum $f^2(||\mathbf{Q}||) + g^2(||\mathbf{Q}||) + f(||\mathbf{Q}||)g(||\mathbf{Q}||)$ is shown for illustrative purposes, highlighting how scattering associated with dipole-dipole correlations can easily mask scattering involving octupoles.}
    \label{Figure7}
\end{figure}

%%%%%%%%%%%%%%%%%%%%%%%%%%%%%%%%%

To further investigate the $Q$-dependence of the measured diffuse scattering, we perform a powder average (an average over direction for each $Q$) for the data in the $(H+0.5,H-0.5,L)$ plane shown in Figures~\ref{Figure3} and \ref{Figure4}. In further detail, for each value of $\|\mathbf{Q}\| = Q$ within range of the measurements, the measured data is averaged over all reciprocal space vectors in the $(H+0.5,H-0.5,L)$ plane with magnitude $Q$. We henceforth refer to this averaging as a powder average over the $(H+0.5,H-0.5,L)$ plane. This averaging results in plots of intensity vs. $Q$ for $\mathbf{Q}$ in the $(H+0.5,H-0.5,L)$ plane, and we show this for the diffraction data itself in Fig.~\ref{Figure5}(a,b) and for the net intensity resulting from a subtraction of the high temperature ($T = 5$~K) data from the low-temperature data in Fig.~\ref{Figure5}(c,d). We show these for three combinations of $Q$-range and intensity (or net intensity) range, which together provide a complete view of the data. 

The full ranges of intensity (or net intensity) and $Q$ are shown in Figs.~\ref{Figure5}(a,c), and the same full intensity range is used with a focus on the low-$Q$ scattering in the insets to Figs.~\ref{Figure5}(a,c). We also show the medium- and high-$Q$ scattering with a narrowed intensity range in Figs.~\ref{Figure5}(b,d), as to allow more detailed scrutiny of the region where any diffuse scattering from magnetic octupoles is expected to be near its maximum ($\sim 8~\angstrom^{-1}$) due to the magnetic form factor for octupoles (see Section~\ref{Sec:IVa}). We have removed data points at the powder Bragg peak locations associated with Bragg scattering from the sample holder and sample environment equipment to give a clean look at the diffuse net scattering from the Ce$_2$Zr$_2$O$_7$ single crystal. 

As can be seen in Fig.~\ref{Figure5}(c), the strongest temperature-dependent diffuse scattering is at relatively low $Q$ (at $Q < 1.5~\angstrom^{-1}$), consistent with the single crystal maps of the $(H+0.5,H-0.5,L)$ plane of reciprocal space shown in Figures~\ref{Figure3} and \ref{Figure4}. Figure~\ref{Figure5}(d) shows that there is \textit{negative} diffuse net scattering observed at somewhat higher $Q$ ($1.5~\angstrom^{-1} \lesssim Q \lesssim 5~\angstrom^{-1}$), but no convincing diffuse net scattering is observed at $Q > 6~\angstrom^{-1}$ where the magnetic form factor for octupolar correlations is predicted to be at a maximum (see Section~\ref{Sec:IVa}).

All of these results for the diffuse net scattering as a function of $Q$ are summarized in Fig.~\ref{Figure6}(a), which shows the absolute value of net-intensity on a logarithmic intensity-scale. Positive diffuse net intensity is plotted in two shades of blue (one for each of the two lowest-temperature datasets), while negative diffuse net intensity is plotted in two different shades of red. Hence, even though the absolute value of the net intensity is employed for these temperature-difference datasets, positive net intensity (the low-temperature intensity is greater than the high-temperature intensity) can be easily distinguished from negative net intensity (the high-temperature intensity is greater than the low-temperature intensity). 

\section{Comparison with Theory}
\label{Sec:IV}
\subsection{Magnetic Form Factors for Dipoles and Octupoles}
\label{Sec:IVa}

Let us first consider the magnetic form factors for neutron scattering from Ce$^{3+}$-based magnetic dipoles and octupoles for the case of $J = 5/2$. An understanding of the relevant magnetic form factors is particularly important when analyzing neutron scattering data, as the scattering from correlations between multipoles has an intensity that scales with the product of the magnetic form factors of the multipoles.

The neutron scattering cross-section, including higher-order multipoles, is usually calculated using the standard hydrogen approximation~\cite{Lovesey1984, Rotter2009, Sibille2020, Smith2022Reply, PlackeThesis}. Here we use the hydrogen approximation with effective charge of $Z_{\mathrm{eff}} = 17$~\cite{Sibille2020}. Figure~\ref{Figure7}(a) shows the $||\mathbf{Q}||$-dependence of the magnetic form factors predicted in Ref.~\cite{PlackeThesis} for both Ce$^{3+}$ dipoles and Ce$^{3+}$ octupoles, for the case of $J = 5/2$ and $m_J = \pm3/2$ (as is relevant for cerium pyrochlores at low temperature~\cite{Gao2019, Gaudet2019, Sibille2020, Poree2022}). We denote these dipolar and octupolar form factors as $f(||\mathbf{Q}||)$ and $g(||\mathbf{Q}||)$, respectively. Specifically, $f(||\mathbf{Q}||)$ and $g(||\mathbf{Q}||)$ correspond to the radial-dependent part of the Fourier transform of the magnetization density for dipoles and octupoles, respectively, re-scaled by any dipole- and octupole-specific constants that arise in the Fourier transform. In fact, the dipole- and octupole-specific constant factors cannot be fully removed from the angular-dependent part of the Fourier transform by simple re-scaling of radial-dependent part, and so the radial-dependent part of the Fourier transform was re-scaled using an approximation that slightly overestimates the size of the octupolar form factor $g(||\mathbf{Q}||)$ compared to the size of the dipolar form factor $f(||\mathbf{Q}||)$. The form factor for dipoles is largest at $||\mathbf{Q}|| = 0$ and decays with increasing $||\mathbf{Q}||$, while the form factor for octupoles is negligible at low $||\mathbf{Q}||$ and exhibits a broad peak centered on $||\mathbf{Q}|| \sim 8~\angstrom^{-1}$.

Figure~\ref{Figure7}(b) shows the relevant products of these form factors for dipoles and octupoles, where $f^2(||\mathbf{Q}||)$ is the product relevant for scattering probing correlations between two dipoles, $g^2(||\mathbf{Q}||)$ is the product relevant for scattering probing correlations between two octupoles, and $f(||\mathbf{Q}||)g(||\mathbf{Q}||)$ is the product relevant for scattering probing correlations between octupoles and dipoles. Figure~\ref{Figure7}(b) also shows the sum of these three products, $f^2(||\mathbf{Q}||) + g^2(||\mathbf{Q}||) + f(||\mathbf{Q}||)g(||\mathbf{Q}||)$, plotted to illustrate how the large intensity of scattering from dipole-dipole correlations can easily mask intensity from correlations involving octupoles. This masking is evidenced by the similar $||\mathbf{Q}||$-dependence of the $f^2(||\mathbf{Q}||)$ (blue) and $f^2(||\mathbf{Q}||) + g^2(||\mathbf{Q}||) + f(||\mathbf{Q}||)g(||\mathbf{Q}||)$ (yellow) curves in Fig.~\ref{Figure7}(b) and the lack of any discernible peaks at nonzero $||\mathbf{Q}||$ in $f^2(||\mathbf{Q}||) + g^2(||\mathbf{Q}||) + f(||\mathbf{Q}||)g(||\mathbf{Q}||)$ (yellow) compared to $g^2(||\mathbf{Q}||)$ (red) and $f(||\mathbf{Q}||)g(||\mathbf{Q}||)$ (green).

\subsection{Numerical-Linked-Cluster Calculations}
\label{Sec:IVb}

We compare our neutron diffraction measurements on single crystal Ce$_2$Zr$_2$O$_7$ to the corresponding predictions according to seventh-order NLC calculations (see Appendix~A for details) using the nearest-neighbor pseudospin interaction parameters in the XYZ Hamiltonian predicted for Ce$_2$Zr$_2$O$_7$ in Ref.~\cite{Smith2022}: $(J_{\tilde{x}},  J_{\tilde{y}},  J_{\tilde{z}})$ = (0.0635, 0.0635, 0.011)~meV and $\theta = 0$. Figure~\ref{Figure6}(b) shows this NLC-calculated temperature-difference neutron diffraction signal powder-averaged over the $(H+0.5,H-0.5,L)$ plane for $T = 0.2$~K with the corresponding $T = 5$~K calculation subtracted. Here, we again use a logarithmic intensity-scale with positive (negative) net intensity shown in blue (red) to provide a like-for-like comparison with the color-coded data from Ce$_2$Zr$_2$O$_7$ in Figure~\ref{Figure6}(a). We also show an overplot of this calculation with the measured data from Ce$_2$Zr$_2$O$_7$ in Figure~\ref{Figure6}(c), using a linear intensity-scale with a reduced maximum intensity to focus on the weaker scattering at $Q \gtrsim 1.5$~\angstrom$^{-1}$. Interestingly, at $Q \gtrsim 1.5$~\angstrom$^{-1}$, both the NLC-calculated signal and the signal measured from Ce$_2$Zr$_2$O$_7$ show oscillations of the net intensity resulting in alternation between positive and negative net intensity with increasing $Q$. These oscillations are clearer in the data measured from Ce$_2$Zr$_2$O$_7$ at low $Q$ ($Q \lesssim 5~\angstrom^{-1}$) compared to at higher $Q$, consistent with the decreasing oscillation size with increasing $Q$ that is predicted by our NLC calculations. 

The NLC calculations describe the location and width of the large positive peak at low $Q$ as well as the smaller positive and negative net intensity peaks resulting from the oscillating net intensity. However, there is room for improvement in this comparison of the measured data with the NLC calculations. For example, the high net-intensity peak at low $Q$ in the measured data is more intense than that predicted by the NLC calculations if the amplitude of the smaller oscillations is to approximately agree between the data and the NLC calculations. Additionally, the positive net intensity peak just below 3~$\angstrom^{-1}$ in the temperature-difference NLC calculations falls in a region where the measured data is plagued by imperfect subtraction of Bragg powder rings associated with scattering from the sample holder and sample environment equipment, and for this reason, no comparison between the calculation and measured data from Ce$_2$Zr$_2$O$_7$ is available in this region. Nonetheless, one can conclude that the diffuse scattering measurements and the NLC calculations using previously-established pseudospin interaction parameters determined for Ce$_2$Zr$_2$O$_7$ are in remarkable qualitative agreement with each other. 

\begin{figure}[t]
\linespread{1}
\par
\includegraphics[width=3.05in]{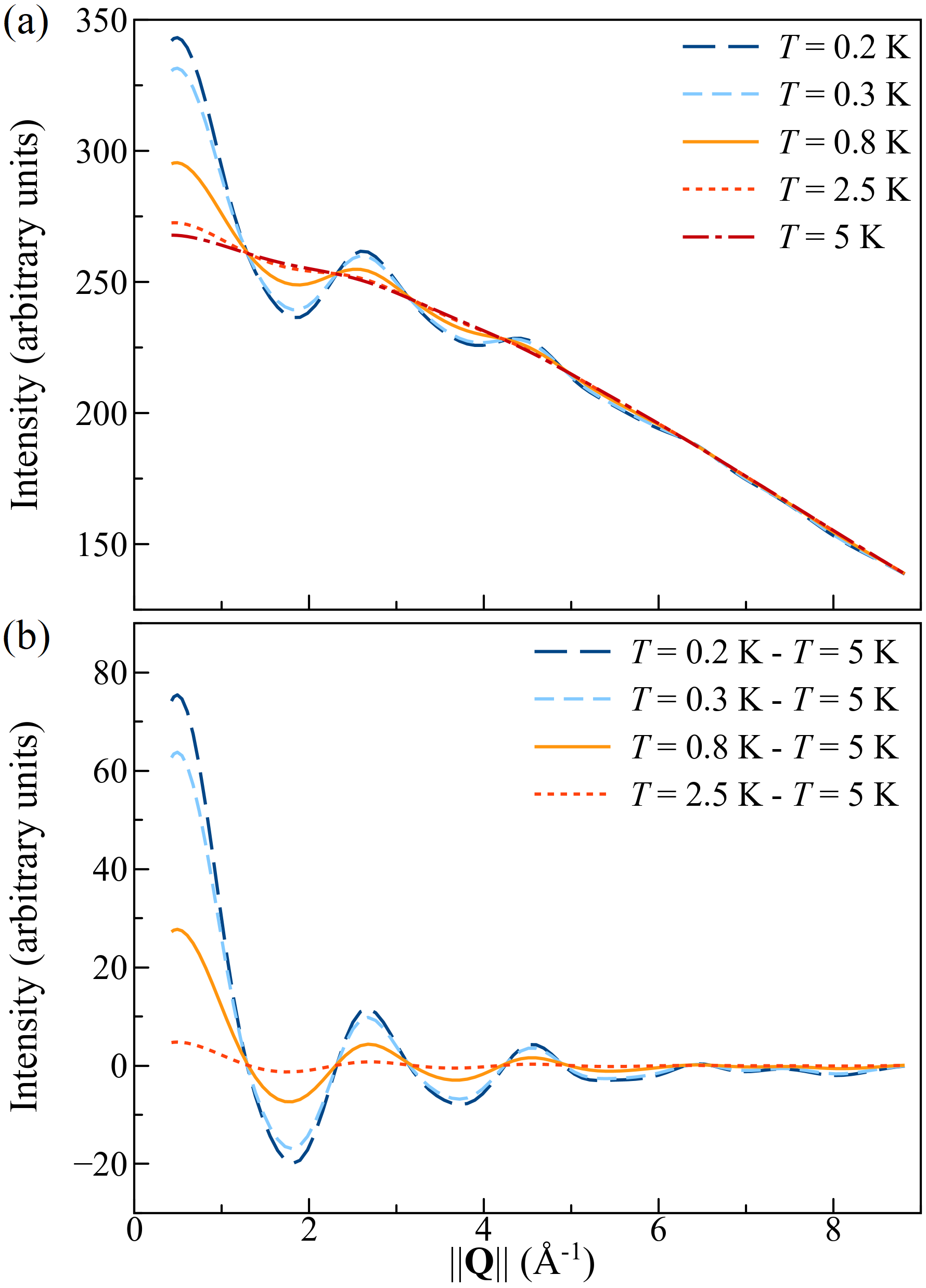}
\par
\caption{(a) Seventh order NLC calculations for the powder averaged neutron diffraction intensity using the parameters for the XYZ Hamiltonian estimated for Ce$_2$Zr$_2$O$_7$ in Ref.~\cite{Smith2022} (see main text), with the powder average taken over the $(H,H,L)$ plane for an integration in $(K,\bar{K},0)$ over $K = [-0.7, -0.3]$ and $K = [0.3, 0.7]$, at various temperatures between $T = 0.2$~K and $T = 5$~K (as labeled). (b) The temperature-difference neutron diffraction signal that is constructed using the calculations shown in (a). Specifically, we show the $T = 0.2$~K, $T = 0.3$~K, $T = 0.8$~K, and $T = 2.5$~K calculations from (a) with the corresponding $T = 5$~K calculation from (a) subtracted in each case.} 
\label{Figure8}
\end{figure}

Fig.~\ref{Figure8}(a) shows the neutron diffraction signal without temperature-subtraction predicted according to these seventh-order NLC calculations for $T = 0.2$~K, $T = 0.3$~K, $T = 0.8$~K, $T = 2.5$~K, and $T = 5$~K. These calculations in Fig.~\ref{Figure8}(a) were used to form the NLC-calculated temperature-difference neutron diffraction shown in Fig.~\ref{Figure6}(b,c), also shown in Fig.~\ref{Figure8}(b) using a wider, linear intensity-scale than that used in Fig.~\ref{Figure6}(c). Importantly, the oscillations in the signal measured from Ce$_2$Zr$_2$O$_7$ are larger for the $T = 0.05$~K$~-~T = 5$~K and the $T = 0.3$~K$~-~T = 5$~K data compared to the $T = 0.8$~K$~-~T = 5$~K data [see Fig.~\ref{Figure5}(d)], consistent with the temperature dependence predicted by our NLC calculations in Fig.~\ref{Figure8}(b). 

%%%%%%%%%%%%%%%%%%%%%%%%%%%%%%%%%
%%%%%%%%%%%%%%%%%%%%%%%%%%%%%%%%%

%%%%%%%%%%%%%%%%%%%%%%%%%%%%%%%%%
\begin{table*}[]
\label{Tab:I}
\begin{tabular}{|c|c|c|c|c|c|c|c|}
\hline
Material & Ref. & $J_{\tilde{x}}$ (meV) & $J_{\tilde{y}}$ (meV) & $J_{\tilde{z}}$ (meV) & $\theta/\pi$ & Ground State & Fig. \\ \hline

\begin{tabular}[c]{@{}c@{}} \end{tabular}          
Ce$_2$Zr$_2$O$_7$ & \cite{Smith2022} & 0.0635 & 0.0635 & 0.011 & 0 & dipolar-octupolar $\pi$-flux QSI & \ref{Figure6}(b,c), \ref{Figure8}, \ref{Figure9}(f) \\ \hline \hline

\begin{tabular}[c]{@{}c@{}} \end{tabular}          
Ce$_2$Sn$_2$O$_7$ & \cite{Sibille2020} & 0 & 0.0414 & 0.0026 & 0 & octupolar $\pi$-flux QSI & \ref{Figure9}(a) \\ \hline

\begin{tabular}[c]{@{}c@{}}  \end{tabular}          
Ce$_2$Sn$_2$O$_7$ & \cite{Yahne2024} & 0.045 & -0.001 & -0.012 & 0.22 & all-in all-out dipole order & \ref{Figure9}(b) \\ \hline \hline

\begin{tabular}[c]{@{}c@{}}  \end{tabular}          
Ce$_2$Hf$_2$O$_7$ & \cite{Poree2023b} & 0.0167 & 0.044 & 0.0103 & 0.1074 & octupolar $\pi$-flux QSI & \ref{Figure9}(c) \\ \hline

\begin{tabular}[c]{@{}c@{}}  \end{tabular}          
Ce$_2$Hf$_2$O$_7$ & \cite{Poree2023b} & 0.0252 & 0.047 & 0.0078 & -0.1844 & octupolar $\pi$-flux QSI & \ref{Figure9}(d) \\ \hline 

\begin{tabular}[c]{@{}c@{}}  \end{tabular}          
Ce$_2$Hf$_2$O$_7$ & \cite{Poree2023b} & 0.046 & 0.022 & 0.011 & -0.0091 & dipolar $\pi$-flux QSI & \ref{Figure9}(e) \\ \hline

\end{tabular}
\caption{Various estimates of $J_{\tilde{x}}$,  $J_{\tilde{y}}$, $J_{\tilde{z}}$, and $\theta$ for the XYZ Hamiltonians [\autoref{Eq:2}] of Ce$_2$Zr$_2$O$_7$, Ce$_2$Sn$_2$O$_7$, and Ce$_2$Hf$_2$O$_7$~\cite{Smith2022, Sibille2020, Poree2023b, Yahne2024}, which we use for the NLC calculations of Figs.~\ref{Figure6}(b,c), \ref{Figure8}, and \ref{Figure9} as labelled in the rightmost column. We show the parameters estimated for Ce$_2$Sn$_2$O$_7$ in Ref.~\cite{Yahne2024} with a slight adjustment to the value of $\theta$ estimated for Ce$_2$Sn$_2$O$_7$ in Ref.~\cite{Yahne2024} (see main text). This table also includes the magnetic ground state phase predicted at the nearest-neighbor level according to the zero-field ground state phase diagram for the XYZ Hamilton [\autoref{Eq:2}] in Ref.~\cite{Benton2020}. The parameters estimated for Ce$_2$Zr$_2$O$_7$ in Ref.~\cite{Smith2022} lie on the boundary between dipolar and octupolar $\pi$-flux QSIs.}
\end{table*}
%%%%%%%%%%%%%%%%%%%%%%%%%%%%%%%%%

%%%%%%%%%%%%%%%%%%%%%%%%%%%%%%%%%
\begin{figure*}[]
\linespread{1}
\par
\includegraphics[width=7in]{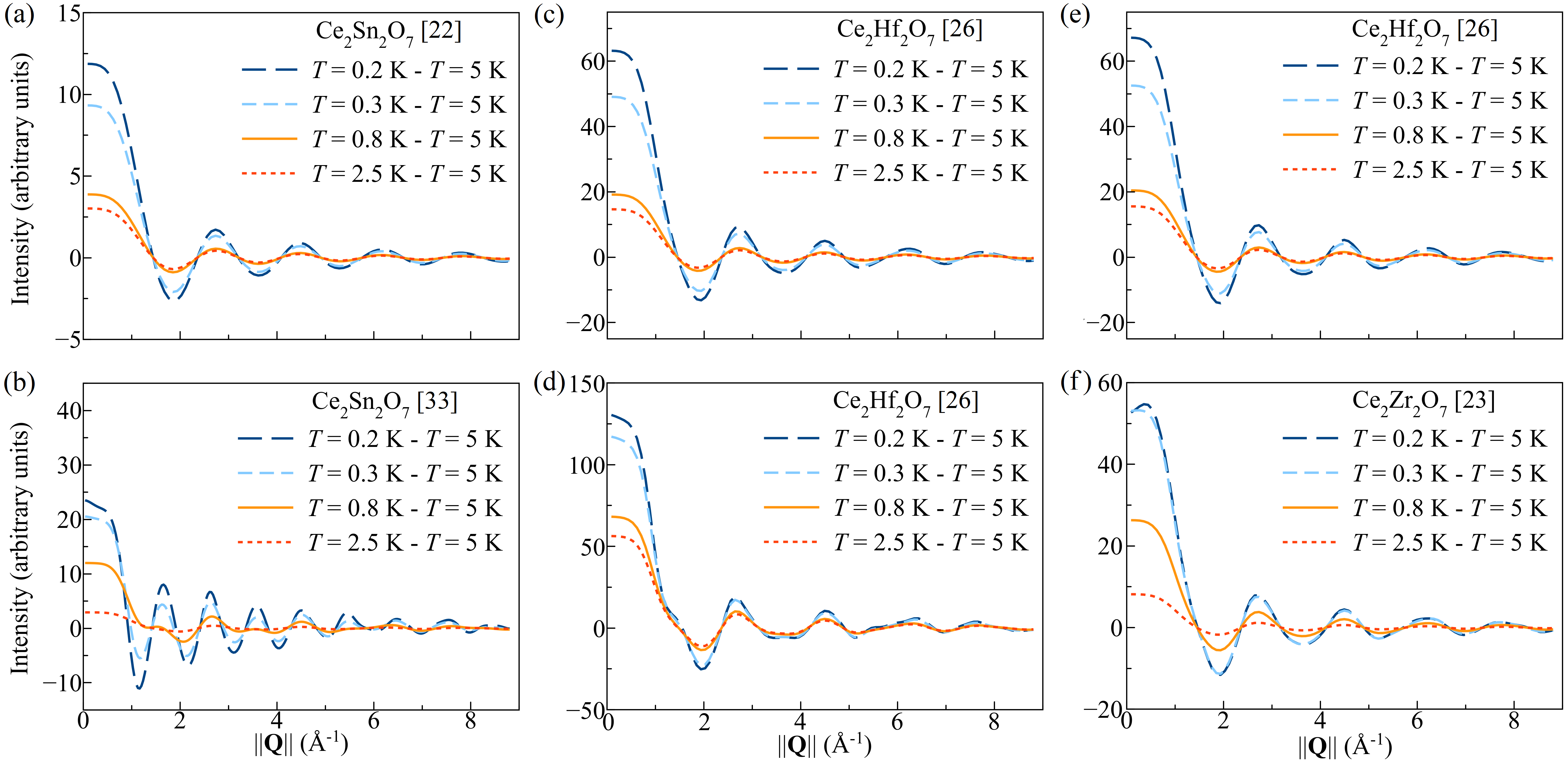}
\par
\caption{Sixth-order NLC calculations for the powder-averaged temperature-difference neutron diffraction signal using various parameters estimated for cerium pyrochlores, shown in each case for $T = 0.2$~K, $T = 0.3$~K, $T = 0.8$~K, and $T = 2.5$~K with the corresponding $T = 5$~K calculation subtracted. Specifically, we use the parameters for the XYZ Hamiltonian suggested for (a) Ce$_2$Sn$_2$O$_7$ in Ref.~\cite{Sibille2020} and (b) for Ce$_2$Sn$_2$O$_7$ in Ref.~\cite{Yahne2024} with $\theta = 0.22 \pi$ slightly adjusted from the value reported in Ref.~\cite{Yahne2024} (see main text). We also show these calculations using the parameters estimated for (c-e) Ce$_2$Hf$_2$O$_7$ in Ref.~\cite{Poree2023b}, and for (f) Ce$_2$Zr$_2$O$_7$ in Ref.~\cite{Smith2022}. These parameters are shown in Table~I.} 
\label{Figure9}
\end{figure*}
%%%%%%%%%%%%%%%%%%%%%%%%%%%%%%%%%

As estimates for the pseudospin Hamiltonian parameters have also been made for Ce$_2$Zr$_2$O$_7$'s sister pyrochlores Ce$_2$Sn$_2$O$_7$ and Ce$_2$Hf$_2$O$_7$, it is also possible to use NLC calculations to examine the theoretically expected diffuse scattering for these materials as represented by these Hamiltonian parameters. Importantly, for each material, these estimates have been arrived at through fitting the XYZ Hamiltonian parameters to the measured heat capacity and the measured magnetization or magnetic susceptibility~\cite{Sibille2020, Smith2022, Poree2023b, Yahne2024}. Accordingly, these calculations represent the respective portions of parameter space for which a good description of the available experimental data can be obtained. Fig.~\ref{Figure9} shows the temperature-difference powder neutron diffraction signal predicted by sixth-order NLC calculations for each of the XYZ Hamiltonian parameter sets suggested in Refs.~\cite{Sibille2020, Poree2023b, Yahne2024}, for  $T = 0.2$~K, $T = 0.3$~K, $T = 0.8$~K, and $T = 2.5$~K with a $T = 5$~K dataset subtracted in each case. Specifically, we use the parameters for the XYZ Hamiltonian suggested for Ce$_2$Sn$_2$O$_7$ in Ref.~\cite{Sibille2020} [Fig.~\ref{Figure9}(a)], a variant of the XYZ Hamiltonian parameters used to describe Ce$_2$Sn$_2$O$_7$ in Ref.~\cite{Yahne2024} [Fig.~\ref{Figure9}(b)], and for three representative XYZ Hamiltonian parameter sets suggested for Ce$_2$Hf$_2$O$_7$ in Ref.~\cite{Poree2023b} [Fig.~\ref{Figure9}(c-e)]. We also show these calculations using the XYZ Hamiltonian parameters estimated for Ce$_2$Zr$_2$O$_7$ in Ref.~\cite{Smith2022} [Fig.~\ref{Figure9}(f)]. These parameter sets are shown in Table~I alongside the corresponding magnetic ground states predicted at the nearest-neighbor level~\cite{Benton2020}.

For each of these parameter sets, the NLC-predicted diffraction signal in Figure~\ref{Figure9} shows a large peak at low $Q$ followed by oscillations of the net intensity, between positive and negative net scattering, with increasing $Q$. As we discuss in the following section, these NLC calculations show qualitative agreement with the corresponding Ce$_2$Sn$_2$O$_7$ data in Ref.~\cite{Yahne2024} and disagree with the corresponding Ce$_2$Sn$_2$O$_7$ and Ce$_2$Hf$_2$O$_7$ diffraction data in Refs.~\cite{Sibille2020, Poree2023b}, which do not show any appreciable diffuse scattering at low $Q$. Interestingly, Refs.~\cite{Sibille2020, Poree2023b} both report temperature-difference \textit{inelastic neutron scattering} data with a signal at low $Q$ that is not detected in the \textit{neutron diffraction} presented in the same work.

The Ce$_2$Sn$_2$O$_7$ parameters we use for the NLC calculations in Fig.~\ref{Figure9}(b) are slightly adjusted from those reported in Ref.~\cite{Yahne2024}, as we use a $\theta$ value of $\theta = 0.22\pi$ rather than the value $\theta = 0.19\pi$ reported in Ref.~\cite{Yahne2024}. This small increase in $\theta$ results in a better description of the neutron diffraction data from Ref.~\cite{Yahne2024} by the NLC calculations, especially with regard to the dominance of the net-positive low-$Q$ diffuse scattering in the high-temperature-subtracted signal measured from Ce$_2$Sn$_2$O$_7$; In contrast to the NLC calculations with $\theta = 0.22\pi$, the NLC calculations with $\theta = 0.19\pi$ yield net-positive diffuse scattering at low $Q$ ($Q \lesssim 1.5 \angstrom^{-1}$) that is weaker than the oscillations in net-scattering at higher $Q$.

The Ce$_2$Hf$_2$O$_7$ parameters we use for the NLC calculations in Fig.~\ref{Figure9}(c,d,e) result from fits to the heat capacity measured from Ce$_2$Hf$_2$O$_7$ in Ref.~\cite{Poree2023b}. Specifically, these fits in Ref.~\cite{Poree2023b} find the best agreement with the measured heat capacity data from Ce$_2$Hf$_2$O$_7$ over a region of parameter space where $\theta$ is small and where $J_{\tilde{y}} > 0$ dominates over $J_{\tilde{x}}$ and $J_{\tilde{z}}$, and over a similar region where $\theta$ is small and where $J_{\tilde{x}} > 0$ dominates. The parameter sets we use for the NLC calculations in Fig.~\ref{Figure9}(c,d,e) arise as representative parameter sets for these best-fitting regions of parameter space in Ref.~\cite{Poree2023b}.

Importantly, for all pseudospin Hamiltonian parameter sets investigated theoretically in this study, the calculated temperature-difference neutron diffraction signal does not show a broad hump at high $Q$ as was reported for Ce$_2$Sn$_2$O$_7$ and Ce$_2$Hf$_2$O$_7$ in Refs.~\cite{Sibille2020, Poree2023b}.   

\section{Discussion}
\label{Sec:V}

Figure~\ref{Figure6}(a) clearly shows the presence of dipolar QSI correlations at low and intermediate $Q$. These appear as positive (blue) diffuse net scattering at low $Q$ (below $\sim 1.5~\angstrom^{-1}$) and as negative (red) diffuse diffuse net scattering at intermediate $Q$ ($1.5~\angstrom^{-1} \lesssim Q \lesssim 5~\angstrom^{-1}$) in this color-coded plot of $|\mathrm{Intensity}|$ vs. $Q$ for the temperature-difference neutron diffraction data. On the other hand, any existing diffuse neutron scattering from {\it octupolar} correlations possesses an intensity below the observation threshold, which is $\sim$ 0.1 $\%$ of the peak in the temperature-difference diffuse scattering due to {\it dipolar} correlations at low $Q$. This is based on the observation that the intensity of the measured diffuse net scattering near the low-$Q$ peak in Fig.~\ref{Figure6}(a) is $\sim 400$, while the average intensity of the measured data points in the $T = 0.05$~K$~-~T = 5$~K and $T = 0.3$~K$~-~T = 5$~K datasets between 6~$\angstrom^{-1}$ and 12~$\angstrom^{-1}$ is 0.02 $\pm$ 0.19.

The temperature-difference neutron diffraction data reported for a powder sample of Ce$_2$Sn$_2$O$_7$ and a powder sample of the classical spin ice Ho$_2$Ti$_2$O$_7$ in Ref.~\cite{Yahne2024} shows a large peak at low $Q$ in each case, centered near $Q = 0.6~\angstrom^{1}$ and followed by an alternation between positive and negative net intensity with increasing $Q$ as observed in this work for Ce$_2$Zr$_2$O$_7$ and predicted by our NLC calculations. On the other hand, the temperature-difference neutron diffraction data reported for powder Ce$_2$Sn$_2$O$_7$ and powder Ce$_2$Hf$_2$O$_7$ in Refs.~\cite{Sibille2020, Poree2023b} shows no large peak at low $Q$ and no oscillations between positive and negative net intensity. Instead, the only clear signal in the neutron diffraction data of Refs.~\cite{Sibille2020, Poree2023b} is a hump of positive net scattering centered on $Q \sim 8~\angstrom^{1}$ with a full width at half maximum of $\sim 5~\angstrom^{1}$. Our sixth order NLC calculations for the powder averaged temperature-difference neutron diffraction signal using the nearest neighbor exchange parameters for Ce$_2$Sn$_2$O$_7$ and Ce$_2$Hf$_2$O$_7$ suggested in Refs.~\cite{Sibille2020, Poree2023b, Yahne2024} [Fig.~\ref{Figure9}(a-e)] are similar to our calculations in Fig.~\ref{Figure6}(b,c), \ref{Figure8}(b), and \ref{Figure9}(f) using the parameters estimated for Ce$_2$Zr$_2$O$_7$. Specifically, each of these calculations shows a large peak at low $Q$ combined with oscillations of the net intensity resulting in alternation between positive and negative net intensity, with no broad hump of positive scattering at high $Q$. Notably, our calculations are in good qualitative agreement with the neutron diffraction data measured on Ce$_2$Sn$_2$O$_7$ in Ref.~\cite{Yahne2024}, and do not describe the neutron diffraction data measured on  Ce$_2$Sn$_2$O$_7$ and Ce$_2$Hf$_2$O$_7$ in Refs.~\cite{Sibille2020, Poree2023b}.

In Refs.~\cite{Sibille2020, Poree2023b}, the temperature-difference neutron diffraction data measured on powder samples of Ce$_2$Sn$_2$O$_7$ and Ce$_2$Hf$_2$O$_7$ is compared with the powder-averaged elastic neutron scattering signal calculated for an octupolar spin ice at absolute zero temperature. This calculated powder-averaged neutron scattering signal for purely octupolar spin ice shows a broad hump of scattering centered on $Q \sim 8~\angstrom^{1}$, which resembles the diffraction data measured on Ce$_2$Sn$_2$O$_7$ and Ce$_2$Hf$_2$O$_7$ in Refs.~\cite{Sibille2020, Poree2023b}, but contrasts with the oscillating temperature-difference neutron diffraction signals predicted using NLC calculations in this work. For that reason, we point out the differences between the quantum NLC calculations in this work and the calculations of Refs.~\cite{Sibille2020, Poree2023b}. First we note that our quantum NLC calculations compute the equal-time neutron scattering structure factor, which includes both elastic and dynamic scattering appropriate for the energy-integrated signal measured in diffraction measurements, while the Monte Carlo calculations of Refs.~\cite{Sibille2020, Poree2023b} compute the elastic neutron scattering structure factor. In addition to this, the quantum NLC calculations of this work should capture the temperature dependence quantitatively, hence allowing for calculation of temperature-difference neutron scattering signals and their comparison to the experimental data to be done with confidence. This is opposed to the zero-temperature Monte Carlo calculations done at the mean-field level in Refs.~\cite{Sibille2020, Poree2023b}, which use an effective Hamiltonian that excludes the terms responsible for the quantum correlations of an octupolar spin ice. The effective Hamiltonian used in Refs.~\cite{Sibille2020, Poree2023b} also dictates that the octupole spin ice phase represented by the calculations is one that shows no correlations between magnetic dipole moments. However, the excitations of an octupolar spin ice would always have some dipolar component~\cite{Smith2025}, with a corresponding magnetic form factor for dipoles that is much larger than that of octupoles (see Fig.~\ref{Figure7}). The same can be said for any dipolar spin ice and for any ordered ground state of the XYZ Hamiltonian with non-zero $\theta$ parameter~\cite{Smith2025}. In our calculations, we incorporate both dipolar and octupolar contributions together using the XYZ Hamiltonian in \autoref{Eq:2}.

The earlier reports of temperature-dependent neutron diffraction from Ce$_2$Sn$_2$O$_7$ and Ce$_2$Hf$_2$O$_7$ in Refs.~\cite{Sibille2020, Poree2023b} did not identify any diffuse dipolar signal at low $Q$~\cite{Sibille2020, Poree2023b}, even though such diffuse scattering is expected to be strong in all these systems, and it has already been identified in separate low-energy inelastic neutron scattering experiments for all three of Ce$_2$Zr$_2$O$_7$, Ce$_2$Sn$_2$O$_7$, and Ce$_2$Hf$_2$O$_7$~\cite{Sibille2020, Gaudet2019, Gao2019, Smith2022, Poree2023b, Yahne2024}. We also note that a single crystal neutron diffraction study on a different sample of Ce$_2$Zr$_2$O$_7$ showed no evidence for temperature-dependent diffuse scattering at any $Q$~\cite{Gao2022}. 

\begin{figure*}[t]
\linespread{1}
\par
\includegraphics[width=5.5in]{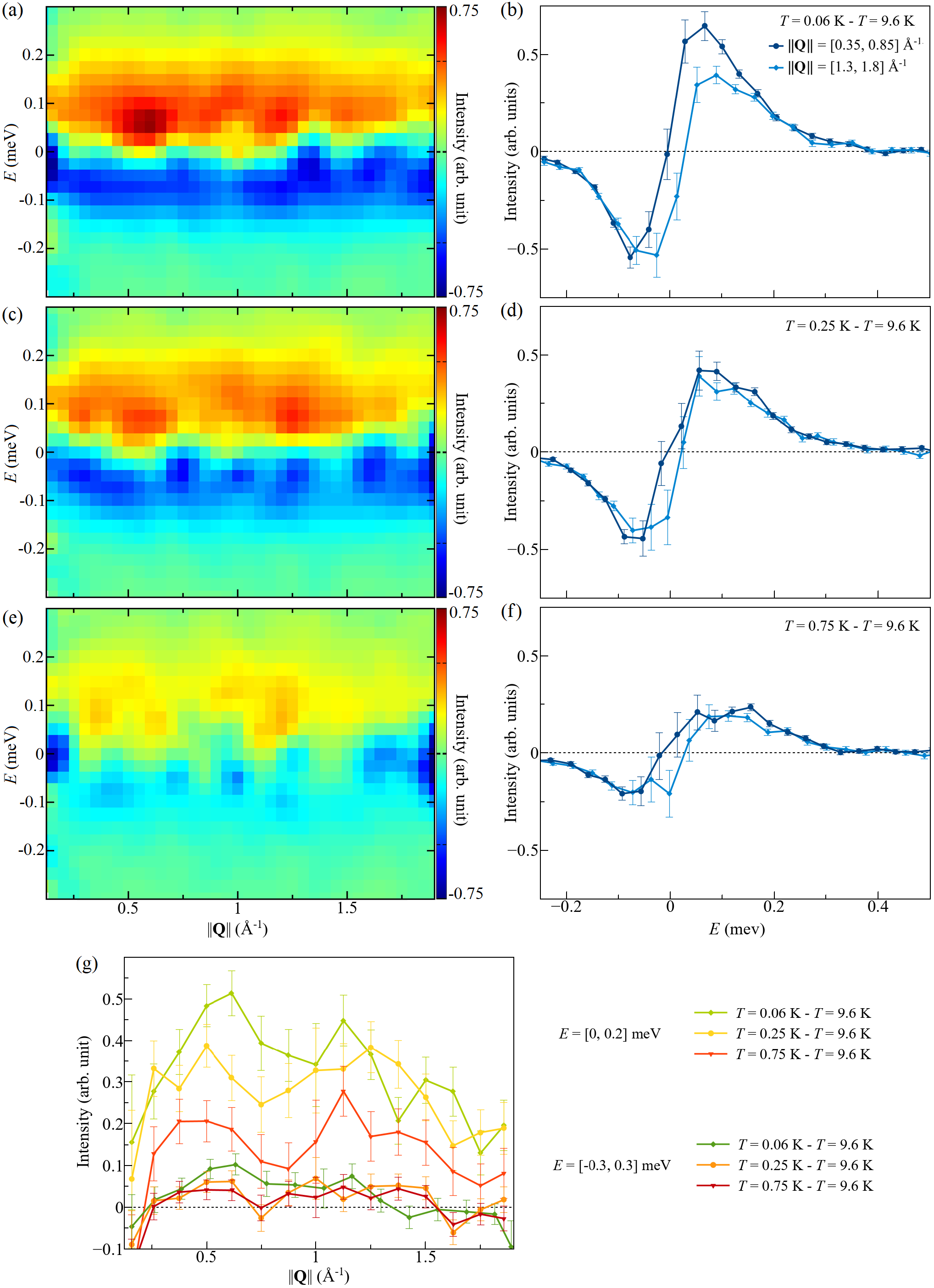}
\par
\caption{The inelastic neutron scattering signal measured (Ref.~\cite{Gaudet2019}) from a powder sample of Ce$_2$Zr$_2$O$_7$ at (a) $T = 0.06$~K, (c) $T = 0.25$~K, and (e) $T = 0.75$~K with a $T = 9.6$~K dataset subtracted in each case. The energy-dependence of the (b) $T = 0.06$~K$~-~T = 9.6$~K, (d) $T = 0.25$~K$~-~T = 9.6$~K, and (f) $T = 0.75$~K$~-~T = 9.6$~K temperature-difference inelastic neutron scattering spectra from a powder sample of Ce$_2$Zr$_2$O$_7$ for $Q$ integration ranges of $[0.35,0.85]~\angstrom^{-1}$ (dark~blue) and $[1.3,1.8]~\angstrom^{-1}$ (light blue)~\cite{Gaudet2019}. The range $[0.35,0.85]~\angstrom^{-1}$ is selected to cover the dominant portion of the measured magnetic signal from Ce$_2$Zr$_2$O$_7$ at low temperature, and both $Q$-integration ranges were chosen to avoid Bragg peak locations. (g) The $Q$-dependence of the $T = 0.06$~K$~-~T = 9.6$~K (green), $T = 0.25$~K$~-~T = 9.6$~K (yellow), and $T = 0.75$~K$~-~T = 9.6$~K (red) temperature-difference inelastic neutron scattering spectra from a powder sample of Ce$_2$Zr$_2$O$_7$. This is shown for energy integrations over the positive energy range $E = [0, 0.2]$~meV (lighter colors), encapsulating the positive net scattering, and over the symmetric energy range $E = [-0.3, 0.3]$~meV, encapsulating both the positive and negative net scattering (darker colors).} 
\label{Figure10}
\end{figure*}

We can use previous inelastic neutron scattering results to understand why this dipolar scattering is harder to observe in diffraction measurements, which integrate in neutron energy transfer ($E$) over the inelastic spectral weight. This analysis is illustrated in Fig.~\ref{Figure10} using low energy inelastic scattering from a powder sample of Ce$_2$Zr$_2$O$_7$; Some of which was previously published in Refs.~\cite{Gaudet2019, Smith2022}. Figure~\ref{Figure10}(a), (c), and (e) show datasets at $T = 0.06$~K, $T = 0.25$~K, and $T = 0.75$~K, respectively, with a high temperature ($T = 9.6$~K) dataset subtracted from each. Figure~\ref{Figure10}(b), (d), and (f), show the energy-dependence of this same temperature-difference inelastic scattering for $Q$ integrations over the ranges $[0.35, 0.85]~\angstrom^{-1}$ and $[1.3,1.8]~\angstrom^{-1}$, which together cover much of the low-$Q$ diffuse scattering while avoiding Bragg peak locations. Figure~\ref{Figure10}(a-f) shows that positive diffuse net scattering is found for positive energies less than $\sim 0.3$~meV only, due in part to the Bose factor, $(1-e^{-E/(k_{\mathrm{B}}T)})^{-1}$, in $S(Q, E) = \chi^{\prime\prime}(Q,E) \times (1-e^{-E/(k_{\mathrm{B}}T)})^{-1}$ where $\chi^{\prime\prime}(Q,E)$ is the imaginary part of dynamic spin susceptibility. On the other hand, negative diffuse net scattering is observed at negative energies greater than $\sim -0.3$~meV.

Integrating this net inelastic scattering over energy transfer symmetrically around $E = 0$~meV, as is necessarily done in a diffraction measurement, results in positive net scattering at low $Q$ in the resulting energy-integrated spectra, but the signal is much diminished for an energy integration that is symmetric around $E = 0$~meV compared to an integration that can distinguish between positive and negative energy transfers to cover only the positive or only the negative net scattering. This is explicitly shown in Fig.~\ref{Figure10}(g), where the energy-integrated signal is shown as a function of $Q$ for an energy integration over $E =[0, 0.2]$~meV, for $T = 0.06$~K$ - T = 9.6$~K, $T = 0.25$~K$ - T = 9.6$~K, and $T = 0.75$~K$ - T = 9.6$~K datasets, compared to the same datasets but integrated in energy over the range $E =[-0.3, 0.3]$~meV. The diffuse low-$Q$ signal in the diffraction data ($E =[-0.3, 0.3]$~meV) is as much as a factor of $\sim 5$ weaker than that with energy discrimination ($E = [0, 0.2]$~meV), and the relative temperature dependence at these low temperatures is also weaker for the diffraction data.

\section{Summary and Conclusions}
\label{Sec:VI}

We have carried out a single crystal diffuse neutron scattering study of the dipole-octupole quantum spin ice candidate pyrochlore Ce$_2$Zr$_2$O$_7$ using the CORELLI diffuse scattering instrument, which provides an energy resolution that is narrower than that of typical neutron diffraction measurements, and which also allows us to extract diffuse scattering from within a large volume of reciprocal space that does not contain any Bragg peaks from single crystal Ce$_2$Zr$_2$O$_7$.  This is important as we use the subtraction of a high-temperature dataset from low-temperature datasets to help isolate the magnetic contribution to the measured scattering. The absence of single-crystal Bragg peaks means that we do not subtract strong Bragg intensities from each other, which are much more intense than the anticipated dipolar and octupolar signals. In addition, we do not have significant contributions from low energy acoustic phonons, which would be captured near Bragg peaks (Brillouin zone centers), and which would not cancel in the subtraction from each other at different temperatures. We conclude that diffuse neutron scattering from octupolar correlations at high $Q$ in Ce$_2$Zr$_2$O$_7$ are not observable at or above the minimum temperature probed in this experiment (0.05~K) and must therefore be below the observation threshold, which is $\sim 0.1 \%$ of the peak in the temperature dependent diffuse scattering due to {\it dipolar} correlations at low $Q$. 

While this is a clear conclusion, it is well-appreciated that Ce$_2$Zr$_2$O$_7$'s pseudospin-$1/2$ degrees of freedom have both dipolar and octupolar character, and there is a very strong case that Ce$_2$Zr$_2$O$_7$'s ground state resides in a $\pi$-flux quantum spin ice phase close to the boundary between dipolar and octupolar character~\cite{Smith2022, Changlani2022, Smith2023}. Accordingly, octupolar correlations are indeed expected at sufficiently low temperatures. However, our calculations of the magnetic form factors for dipoles and octupoles show that the decay with increasing $Q$ in the squared form factor for dipoles can easily mask the relatively small peak at $Q\sim8~\angstrom^{-1}$ in the squared form factor for octupoles. In association with this, our NLC calculations show that correlations from magnetic octupoles are well-hidden in temperature subtractions of neutron diffraction data, such that they do not generate a significant increase in intensity near $Q\sim8~\angstrom^{-1}$ and are instead concealed by the oscillating net intensity that results from the subtraction of dipole-based scattering. 

The dominant signal present in both the measured and NLC-calculated temperature-difference neutron diffraction is positive diffuse net scattering present at small $Q < 1.5~\angstrom^{-1}$ and is attributable to dipolar correlations in the spin ice state at these temperatures~\cite{Gaudet2019, Smith2022, Changlani2022}. Clear but weak negative diffuse net scattering in the measured data follows at intermediate wavevectors, between $\sim 1.5~\angstrom^{-1}$ and $\sim 5~\angstrom^{-1}$, and this is associated with the subtraction of the paramagnetic diffuse scattering at high temperatures ($T = 5$~K) from the diffuse spin ice like scattering at low temperatures. The negative diffuse scattering in this $Q$-region, and the weakening of scattering for $Q$ above this region, are both qualitatively consistent with the net intensity predicted by our NLC calculations. 

Our NLC calculations using the nearest-neighbor pseudospin interaction parameters suggested for Ce$_2$Sn$_2$O$_7$ and Ce$_2$Hf$_2$O$_7$ in Refs.~\cite{Sibille2020, Poree2023b, Yahne2024} are similar to our calculations for Ce$_2$Zr$_2$O$_7$ in that the dominant signal in the temperature-difference neutron diffraction is a peak at low $Q$ and this peak is followed by weaker oscillations between positive and negative net intensity with increasing $Q$. The features in the NLC-calculated temperature-difference neutron diffraction are qualitatively consistent with the data measured from hydrothermally-grown powder samples of Ce$_2$Sn$_2$O$_7$ in Ref.~\cite{Yahne2024}. They are largely inconsistent with the data reported in Refs.~\cite{Sibille2020, Poree2023b} on powders of Ce$_2$Sn$_2$O$_7$ and Ce$_2$Hf$_2$O$_7$ grown through solid state synthesis methods, where only a broad hump of positive net scattering centered on $Q \sim 8~\angstrom^{-1}$ was reported in the temperature-difference neutron diffraction and was attributed to octupolar correlations. Indeed, the calculations of the present work show that octupolar correlations are not likely to generate a large, broad hump at high $Q$ in such measurements.  

This study emphasizes the power of new neutron instrumentation dedicated to diffuse scattering and sheds light on the absence of clear octupolar correlations in the low-temperature neutron signal from Ce$_2$Zr$_2$O$_7$. We hope that our work in understanding this absence highlights the need for experiments on Ce$_2$Zr$_2$O$_7$ and other cerium pyrochlores using probes with significant sensitivity to octupolar correlations. We also hope that our present work can motivate experimental and theoretical work on this forefront problem in quantum materials. 

%%%%%%%%%%%%%%%%%%%%%%%%%%%%%%%%%

%%%%%%%%%%%%%%%%%%%%%%%%%%%%%%%%%

\begin{acknowledgments}
This work was supported by the Natural Sciences and Engineering Research Council of Canada (NSERC). We sincerely thank Jonathan Gaudet and Nicholas Butch for their work in previously-published inelastic neutron scattering measurements on powder Ce$_2$Zr$_2$O$_7$ (Ref.~\cite{Gaudet2019, Smith2022}), which we have used again here with new analysis relevant to this work. We greatly appreciate the technical support from Marek~Kiela and Jim~Garrett at the Brockhouse Institute for Materials Research, McMaster University. We thank the Max Planck Institute for the Physics of Complex Systems for their computational support. A portion of this research used resources at the Spallation Neutron Source, a DOE Office of Science User Facility operated by the Oak Ridge National Laboratory. RS acknowledges the AFOSR Grant No. FA 9550-20-1-0235. 
\end{acknowledgments}

\section*{Appendix A: Details of Numerical-Linked-Cluster Calculations}
 
The numerical-linked-cluster (NLC) expansion has proven a valuable tool to compute equilibrium properties in a broad class of quantum systems~\cite{rigol_nlce_square_2007,rigol_nlce_kagome_square_tri_2007,khatami_nlce_pinwheel_kagome_2011,khatami_nlce_checkerboard_2011,tang_nlce_2013}.
Its success is particularly striking for three-dimensional frustrated systems like the pyrochlore antiferromagnet, where traditional methods like exact diagonalization, quantum Monte Carlo, or tensor-based approaches are severely limited, either due to the three-dimensional nature of the problem or the sign problem induced by the geometrical frustration.
As demonstrated in Refs.~\cite{Schafer2020,schaefer_magnetic_2022}, it allows for controlled and unbiased predictions down to non-trivial temperatures far beyond conventional high-temperature expansions. 
In particular, it has been previously utilized to characterize Cerium-based pyrochlore materials and determine the strengths of nearest neighbor exchange parameters~\cite{Smith2022,Smith2023,Beare2023,Yahne2024}.

Unlike most earlier approaches, we go beyond the traditional application of NLC expansions where it is used to compute extensive thermodynamic quantities like specific heat or magnetic susceptibility.
We have set up a systematic way to obtain spatially resolved observables allowing us to compute the structure factor for a direct comparison to the experiment.
While the computational overhead is only slightly increased, the required framework is much more evolved compared to previous works, and a detailed explanation of the algorithm can be found in Ref.~\cite{schaefer_NLCE_corr_2024}.
Here, we present a brief, but precise, description of the algorithm and point out the main differences compared to the non-spatially-resolved NLC expansion.

\paragraph{Cluster Growth:} The first step is identical to our previous works, and we successively grow the cluster in a systematic way such that we find all connected and translationally invariant subclusters made of tetrahedra.
The centers of the up- and down-pointing tetrahedra lay on the A and B sites of a diamond lattice, respectively, which form the basis of our expansion. Then, we start with two single-site clusters on a single A and B site in the diamond lattice. Then, we systematically add neighboring sites such that the clusters successively grow. Within this process, we remove translationally equivalent clusters. For each order $n$, we obtain a collection of graphs (i.e., clusters) that we denote by $\mathcal{C}_n$.

\paragraph{Topological Reduction:} While the created clusters in $\mathcal{C}_n$ are translationally invariant, their topological structure is often equivalent. This allows us to drastically reduce the computational complexity as only a small fraction of these have to be solved with exact diagonalization. 
All topologically inequivalent clusters of order $n$ are collected in $\mathcal{T}_n$.
For each cluster $c\in\mathcal{C}_n$, there is an isomorphism $\mathbf{I}_c$ mapping it to a cluster $t\in\mathcal{T}_n$. 
Table 9 in Ref.~\cite{Schafer2020} summarizes the number of translational and topologically inequivalent clusters.
For example, for the 9th order (cluster made up of nine tetrahedra), we find $\vert\mathcal{C}_9\vert=79426$ translationally invariant clusters from the first step (paragraph~\textit{a} in this section) that are reduced to $\vert \mathcal{T}_9\vert=49$ topologically invariant clusters.

In the traditional formulation of the NLC expansion without spatial resolution, the large set of connected cluster $\mathcal{C}_n$ can be discarded, and each topologically inequivalent cluster $t\in\mathcal{T}_n$ is associated with multiplicity $L_t\in\mathbb{N}$ that counts the number of clusters $c\in\mathcal{C}_n$ that are mapped to $t$.
The contribution to an extensive, spatially independent observable, such as energy or specific heat, from $t$ is then simply multiplied by $L_t$ to take all connected clusters into account.
For example, if we want to compute the energy at temperature $T$, the contribution of two topologically equivalent clusters $c,c^\prime\in\mathcal{C}_n$ is identical, and we can simply compute it for one cluster and multiply the result by two.

This is, however, not true for spatially resolved observables that depend on the difference between two points $O(\vec{r}-\vec{r^\prime})$ or sites $O(i,j)$ as needed to compute correlation functions.
Instead of only taking the group of topologically inequivalent clusters $\mathcal{T}_n$ with their multiplicities into account, we have included each connected cluster in $\mathcal{C}_n$ individually with its local coordinates.
Fortunately, we do not have to solve each cluster in $\mathcal{C}_n$, which is not feasible in our case, as we can use the solution of the topologically inequivalent clusters in $\mathcal{T}_n$ together with the isomorphisms providing a mapping from $\mathcal{C}_n$ to $\mathcal{T}_n$.
If we have computed the correlation function $O_t(i,j,T)$ between sites $i$ and $j$ for $t\in\mathcal{T}_n$, we can use the inverse of the isomorphism $\mathbf{I}_c$ to obtain the correlation function 
\begin{align}
    O_c(\mathbf{I}_c^{-1}(i),\mathbf{I}_c^{-1}(j),T)=O_t(i,j,T)
\end{align} 
for the connected cluster $c\in\mathcal{C}_n$.

\paragraph{Computing the Observables:} This part is mostly equivalent to the standard NLC algorithm, but here, we additionally compute all two-point correlation functions $O_t(i,j,T)$ for each topologically invariant cluster $t\in\mathcal{T}_n$.
This increases the computational complexity only slightly, as eigenvectors are additionally required.

\paragraph{Computing the Weights:} The computation of the weights of the two-point correlation function is much more involved compared to non-spatially-resolved observables. To determine the weight of a topologically inequivalent graph $t\in\mathcal{T}_n$, we identify all subgraph isomorphisms for all topologically inequivalent clusters $p\in\mathcal{T}_m$ of lower order $m<n$. 
A subgraph isomorphism defines an equivalence between a smaller graph $p$ with a subgraph of a larger graph $t$.
We denote the collection of all subgraph isomorphisms between $p$ and $t$ by $\mathcal{S}(p,t)$.
For example, in the case of the pyrochlore lattice, there are two subgraph isomorphisms that embed a cluster given by two adjacent tetrahedra into a cluster of three adjacent tetrahedra.

The associated weight $W_t(i,j,T)$ of an observable $O_t(i,j,T)$ (like the $S^zS^z$ correlation between sites $i$ and $j$) is obtained using the following formula:
\begin{align}
   & W_t(i,j,T) = O_t(i,j,T)&&\label{eq:weights}\\
   &\quad - \sum_{m<n}~\sum_{p\in\mathcal{T}_m}~\sum_{\mathbf{I}\in\mathcal{S}(p,t)}\sum_{k,l\in p}\delta_{\mathbf{I}(k),i}\delta_{\mathbf{I}(l),j}W_p(k,l,T)\nonumber .
\end{align}
The first and second sums take all topologically inequivalent clusters of a smaller order, i.e., containing fewer tetrahedra than $t$, into account.
The third sum considers all subgraph isomorphisms embedding cluster $p$ into $t$.
The delta functions ensure that sites $k$ and $l$ of cluster $p$ are mapped to sites $i$ and $j$ of cluster $t$, respectively.

\begin{figure}[t]
\linespread{1}
\par
\includegraphics[width=3.4in]{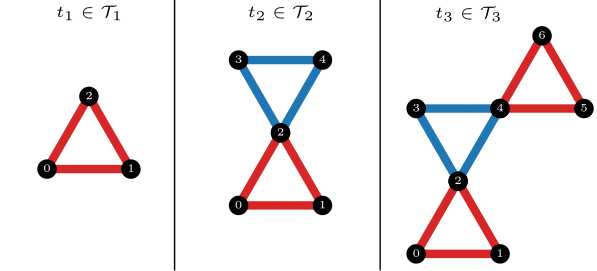}
\par
\caption{Illustration of the numerical-linked-cluster expansion for the kagome lattice. $t_i\in\mathcal{T}_i$ are all the topologically inequivalent clusters up to order three in the triangular expansion. Red and blue triangles are the two building blocks and correspond to the up- and down-pointing triangles. Note that the triangular expansion in the kagome lattice is the two-dimensional version of the tetrahedron expansion in the pyrochlore lattice and is used here for illustrative purposes.} 
\label{Figure11}
\end{figure}

To clarify the rather technical formulation of \autoref{eq:weights}, we outline the procedure for the concrete example in Fig.~\ref{Figure11}.
For illustration purposes, we used the triangular expansion for the kagome lattice, which is the two-dimensional equivalent to the three-dimensional tetrahedral expansion for pyrochlores.
We start by evaluating the first order defined by the cluster $t_1$ where all nearest neighbor weights are given by the observable itself:
\begin{align}
    W_{t_1}(i,j,T) = O_{t_1}(i,j,T)\,.
\end{align}
To obtain the second-order contributions $ W_{t_2}(i,j,T)$, we need to subtract the first-order weights.
There are two subgraph isomorphisms that embed $t_1$ into two subgraphs of $t_2$ spanned by sites $\left\{ 0,1,2\right\}$ and $\left\{ 2,3,4\right\}$:
\begin{align}
    \mathbf{I}_{1\rightarrow 2,a}: \left\{\begin{array}{lr}
        0\mapsto 0\\
        1\mapsto 1\\
        2\mapsto 2
        \end{array}\right.,\,
        \mathbf{I}_{1\rightarrow 2,b}: \left\{\begin{array}{lr}
        0\mapsto 2\\
        1\mapsto 3\\
        2\mapsto 4
        \end{array}\right.\nonumber .
\end{align}
If we want to compute the correlation between site $2$ and $4$, we find that only the second isomorphism, $\mathbf{I}_{1\rightarrow 2,b}$, applies and \autoref{eq:weights} then gives:
\begin{align}
    W_{t_2}(2,4,T) = &\,O_{t_2}(2,4,T)\\
    &\,- W_{t_1}(\mathbf{I}_{1\rightarrow 2,b}^{-1}(2),\mathbf{I}_{1\rightarrow 2,b}^{-1}(4),T)\nonumber\\
    =&\,O_{t_2}(2,4,T)-O_{t_1}(0,2,T)\nonumber .
\end{align}
This procedure is further extended to the third order, where we compute the weight $W_{t_3}(2,4,T)$.
Similarly, we find three subgraph isomorphisms mapping the triangle $t_1$ into the three adjacent triangles of $t_3$:
\begin{align}
    \mathbf{I}_{1\rightarrow 3,a}: \left\{\begin{array}{lr}
        0\mapsto 0\\
        1\mapsto 1\\
        2\mapsto 2
        \end{array}\right.,\,
        \mathbf{I}_{1\rightarrow 3,b}: \left\{\begin{array}{lr}
        0\mapsto 2\\
        1\mapsto 3\\
        2\mapsto 4
        \end{array}\right., \,
        \mathbf{I}_{1\rightarrow 3,c}:\left\{\begin{array}{lr}
        0\mapsto 4\\
        1\mapsto 5\\
        2\mapsto 6
        \end{array}\right.\nonumber .
\end{align}
There are only two subgraph isomorphisms that map $t_2$ to onto $t_3$:
\begin{align}
    \mathbf{I}_{2\rightarrow 3,a}: \left\{\begin{array}{lr}
        0\mapsto 0\\
        1\mapsto 1\\
        2\mapsto 2\\
        3\mapsto 3\\
        4\mapsto 4
        \end{array}\right.,\,
        \mathbf{I}_{2\rightarrow 3,b}: \left\{\begin{array}{lr}
        0\mapsto 2\\
        1\mapsto 3\\
        2\mapsto 4 \\
        4\mapsto 5\\
        5\mapsto 6
        \end{array}\right.\label{eq:iso_1to3} .
\end{align}
By identifying the correct bonds, we arrive at:
\begin{align}
    W_{t_3}(2,4,T) = &\,O_{t_3}(2,4,T)\\
    &\,- W_{t_1}(\mathbf{I}_{1\rightarrow 3,b}^{-1}(2),\mathbf{I}_{1\rightarrow 3,b}^{-1}(4),T)\nonumber\\
    &\,- W_{t_2}(\mathbf{I}_{2\rightarrow 3,a}^{-1}(2),\mathbf{I}_{2\rightarrow 3,a}^{-1}(4),T)\nonumber\\
    &\,- W_{t_2}(\mathbf{I}_{2\rightarrow 3,b}^{-1}(2),\mathbf{I}_{2\rightarrow 3,b}^{-1}(4),T)\nonumber\\
    = &\,O_{t_3}(2,4,T)- W_{t_1}(0,2,T)\nonumber\\
    &\,- W_{t_2}(2,4,T)- W_{t_2}(0,2,T)\nonumber .
\end{align}
Note that we obtain a different result if we compute $W_{t_3}(5,6,T)$ instead.
Since, the bond $(5,6)$ (see Fig.~\ref{Figure11}) occurs only once in the subgraph isomorphisms of \autoref{eq:iso_1to3}, we only find a single contribution of $t_2$ to this bond:
\begin{align}
    W_{t_3}(5,6,T) = &\,O_{t_3}(5,6,T)\\
    &\,- W_{t_1}(\mathbf{I}_{1\rightarrow 3,c}^{-1}(5),\mathbf{I}_{1\rightarrow 3,c}^{-1}(6),T)\nonumber\\
    &\,- W_{t_2}(\mathbf{I}_{2\rightarrow 3,b}^{-1}(5),\mathbf{I}_{2\rightarrow 3,b}^{-1}(6),T)\nonumber\\
    = &\,O_{t_3}(5,6,T)- W_{t_1}(1,2,T)- W_{t_2}(4,5,T)\nonumber .
\end{align}

%%%%%%%%%%%%%%%%%%%%%%%%%%%%%%%%%
%%%%%%%%%%%%%%%%%%%%%%%%%%%%%%%%%

\paragraph{Final Summation:} 
As discussed in the preceding paragraphs, each connected cluster $c\in\mathcal{C}_n$ contributes in our case.
To determine the contribution of $c$ we use the weights computed for its isomorphic cluster $t\in\mathcal{T}_n$ together with its isomorphism $\mathbf{I}_c$.
In the end, the NLC method provides an estimate for $O(\vec{r},T)$ where $\vec{r}$ is the vector between two points in the lattice.
Each connected cluster has a list of coordinates $\vec{r}_i$ associated with each site $i$.
If the vector between two sites, $i$ and $j$, matches the desired $\vec{r}=\vec{r}_i-\vec{r}_j$, they contribute to the expectation value with their weight $W_c(i,j,T)$.
Note that the weight is only computed for $t$ and not for $c$, but can be extracted using the isomorphism $\mathbf{I}_c$ mapping $c$ to $t$:
\begin{align}
    W_c(i,j,T) = W_t(\mathbf{I}_c^{-1}(i),\mathbf{I}_c^{-1}(j),T)\label{eq:iso_ctot} .
\end{align}
The total contribution to $O_m(\vec{r},T)$ up to order $m$ is given by:
\begin{align}
    O_m(\vec{r},T) =& \sum_{n<m}\sum_{c\in\mathcal{C}_n}\sum_{i,j\in c}\delta_{\vec{r}_i-\vec{r}_j,\vec{r}}W_c(i,j,T)\label{eq:nlce_sum}\\
    =& \sum_{n<m}\sum_{c\in\mathcal{C}_n}\sum_{i,j\in c}\delta_{\vec{r}_i-\vec{r}_j,\vec{r}}W_t(\mathbf{I}_c^{-1}(i),\mathbf{I}_c^{-1}(j),T)\nonumber .
\end{align}
We have used \autoref{eq:iso_ctot} to extract the weight of cluster $c$ from its topologically equivalent cluster $t$.
The final expectation value is normalized by the number of sites, and it approximates:
\begin{align}
    O(\vec{r},T) =& \lim_{m\rightarrow\infty}O_m(\vec{r},T)\\
    =&\lim_{N\rightarrow\infty}\frac{1}{N}\sum_{i,j=0}^{N-1} \delta_{\vec{r}_i-\vec{r}_j,\vec{r}}O(i,j,T)\nonumber .
\end{align}

\paragraph{Computing the Structure Factor:} 
The algorithm provides a finite temperature estimate for correlation functions $O_m(\vec{r},T)$ up to a distance given by $m$ expansion units. 
In the case of the tetrahedra in the pyrochlore lattice (or triangles in the kagome lattice), this corresponds to $m$ lattice sites.
The Fourier transformation of all computed expectation values is simply obtained by
\begin{align} \label{eq:nlce_final}
    O_m(\vec{Q},T) = \sum_{\vec{r}\in\mathcal{R}} e^{i \vec{Q}\cdot\vec{r}}O_m(\vec{r},T)\,,
\end{align}
where $\mathcal{R}$ is a collection of all possible $\vec{r}$ values up to order $m$. In contrast to simulations on finite lattices, the NLC expansion computes the expectation value on an infinite lattice using translational symmetry.
Therefore, using the above formula, the grid of $\vec{Q}$-space can be arbitrarily fine.

\paragraph{Dipolar-Octupolar Structure Factor:}

To compute the neutron scattering cross section for dipolar-octupolar pyrochlores we preceded similarly to Ref.~\cite{Yahne2024} where the pseudospin correlations are obtained by classical Monte Carlo simulation.
Now, with the NLC algorithm, we have an unbiased and controlled way to extend this into the quantum regime.
To do so, we have to use \autoref{eq:nlce_sum} and relate the correlations of pseudospin operators to scattering amplitudes. 

To this end, one first has to distinguish correlations between pseudospins residing on the four different rare-earth sublattices composing the face-centered cubic network of $R^{3+}$ tetrahedra in the $R_2B_2$O$_7$ pyrochlores~\cite{Ross2011, Huang2014}.
We then use \autoref{eq:nlce_final}, but keep track of both the sublattice indices $i,j\in\{0,1,2,3\}$ and the components of the pseudospin operators $\alpha,\beta\in\{x,y,z\}$, to compute the correlation matrix
\begin{align}
    \mathcal C_{ij}^{\alpha\beta}(\vec{Q},T) =& \frac{1}{N}\sum_{\vec{r}, \vec{r}^\prime} \exp[-\mathrm{i}\vec{Q} \cdot (\vec{r} - \vec{r}^\prime)]\nonumber\\
    &\times  \expval{S_i^\alpha(\vec{r},T) S_j^\beta(\vec{r}^\prime,T)} \; 
\end{align} 
Similar to the standard NLC formulation, higher orders improve the convergence, yielding an estimate for the structure factor in the thermodynamic limit at lower temperatures. 
Ref.~\cite{schaefer_NLCE_corr_2024} provides a detailed analysis regarding the convergence and limitations.
Using the sublattice- and component-resolved correlations, we compute the equal-time neutron-scattering cross-section,
\begin{align}
&S(\mathbf{Q}, T) = \sum_{ij} 
\exp[-\mathrm{i}\vec{Q} \cdot (\vec{x}_i - \vec{x}_j)] \nonumber\\
&~\times \sum_{\alpha, \beta, \delta, \gamma}
\left(\delta_{\gamma\delta} - \frac{Q^{\gamma}Q^{\delta}}{\abs{\mathbf{Q}}^2}\right) \mathcal D_i^{\alpha\gamma}(\mathbf{Q})D_i^{\beta\delta}(\mathbf{Q}) \mathcal C_{ij}^{\alpha\beta}(\mathbf{Q}, T)
\end{align}
where $\mathbf x_i$ is the position of sublattice $i$ and $\mathcal D_i(\mathbf Q)$ is the momentum-dependent form factor on sublattice $i$, which is given explicitly in Appendix F of Ref.~\cite{Yahne2024} for example (see also e.g. Appendix D of Ref.~\cite{PlackeThesis} for a detailed derivation).

For the calculations shown in Figs.~\ref{Figure6}-\ref{Figure8}, which we compare to our single-crystal diffraction data, $S(\vec{Q},T)$ is calculated for $\mathbf{Q} = (H+K,H-K,L)$ and subsequently averaged for each $H$ and $L$ over $K$ in the ranges $K = [-0.7, -0.3]$ and $K = [0.3, 0.7]$. This averaging over $K$ results in the same $(K,\bar{K},0)$ integration used for the measured data (Figs.~\ref{Figure3}-~\ref{Figure6}). We then perform a powder average over the $(H+0.5,H-0.5,L)$ plane for this integrated $S(\vec{Q},T)$, as described in Section~\ref{Sec:III} for the measured data. For the calculations shown in Fig.~\ref{Figure9}, we perform a full powder average over all directions to allow like-for-like comparison with the powder diffraction data measured from Ce$_2$Sn$_2$O$_7$ and Ce$_2$Hf$_2$O$_7$ in Refs.~\cite{Sibille2020, Poree2023b, Yahne2024}. We perform these NLC calculations at various temperatures between $T = 0.2$~K and $T = 5$~K. The lowest temperature used in the NLC calculations is $T = 0.2$~K due to the fact that the NLC calculations have a low-temperature cutoff where the accuracy of the calculations cannot be verified below the cutoff (further details in Refs.~\cite{Schafer2020,schaefer_magnetic_2022, Smith2023, Yahne2024, schaefer_NLCE_corr_2024} for example). The cutoff temperature decreases with increasing order of the NLC calculations, and also depends on the interaction parameters used the calculations. This cutoff temperature is $\lesssim$ 0.2 K for each of the NLC calculations used in this work.

%%%%%%%%%%%%%%%%%%%%%%%%%%%%%%%%%
\bibliography{Main.bib}
%%%%%%%%%%%%%%%%%%%%%%%%%%%%%%%%%
\end{document}